\documentclass[pdflatex,sn-aps]{sn-jnl}

\usepackage{graphicx}
\usepackage{amsmath,amssymb,amsfonts}
\usepackage{booktabs}
\usepackage{array}
\usepackage{tabularx}
\newcolumntype{Y}{>{\raggedright\arraybackslash}X}
\usepackage{xcolor}
\let\orcidlogo\undefined 
\usepackage{orcidlink}
\usepackage[title]{appendix}

\newcommand{\Smax}{S_{\mathrm{vN}}^{\max}}
\newcommand{\chimax}{\chi_{99}^{\max}}
\newcommand{\Mmax}{M_{2}^{\max}}
\providecommand{\openone}{\mathbb{I}}          
\graphicspath{{figs/}}
\hypersetup{bookmarksdepth=3} 

\usepackage{changes}

\begin{document}

\title[Comparing classical and quantum QUBO solvers]{Comparing classical
and quantum QUBO solvers: optimality, trajectory diagnostics, and limits
of solver selection}

\author*[1]{\fnm{Gerhard} \sur{Hellstern}\orcidlink{0000-0002-6857-6634}}
\email{gerhard.hellstern@dhbw-stuttgart.de}

\author[1]{\fnm{Cedric} \sur{Schacht}\orcidlink{0009-0001-2963-7326}}
\email{cedric.schacht@dhbw-stuttgart.de}

\author[2]{\fnm{Martin} \sur{Zaefferer}\orcidlink{0000-0003-2372-2092}}
\email{zaefferer@dhbw-ravensburg.de}

\affil*[1]{\orgname{Duale Hochschule Baden-W\"urttemberg (DHBW)
Stuttgart}, \orgaddress{\city{Stuttgart}, \country{Germany}}}

\affil[2]{\orgname{Duale Hochschule Baden-W\"urttemberg (DHBW)
Ravensburg}, \orgaddress{\city{Ravensburg}, \country{Germany}}}

\abstract{Comparing solvers for quadratic unconstrained binary optimization
(QUBO) requires distinguishing solution quality from optimality
certification. We compare classical, tensor-network, and fixed-schedule
quantum approaches on 180 benchmark instances from ten families with 14--48
variables, supplemented by harder synthetic instances with known optimal
solutions, extending to 128 variables. On the main benchmark, a general-purpose
branch-and-cut solver matches all reference energies and certifies 176
instances within a 600\,s cap. When the known optima are moved to randomly
chosen bitstrings, the search comparison changes: at $n = 48$ the
branch-and-cut solver finds only four of ten optima and
certifies none, whereas parallel tempering reaches every optimum in every
tested run at a 60\,s budget. Structural and trajectory diagnostics
explain aspects of tensor-network performance. A treewidth bound ranks
trajectory bond-dimension demand at Spearman $\rho = 0.78$ in sample and
$0.73$ out of sample, but provides only a coarse runtime estimate. Controlled
experiments identify excited-state trapping during warm-started
density-matrix renormalization group (DMRG) sweeps and demonstrate mitigation by restarting the variational search.
These diagnostics do not yield a validated solver-selection rule.
The tested linear-ramp quantum approximate optimization algorithm (LR-QAOA) runs offer no solution-quality improvement
over the classical comparators. In the separate Wishart scaling study, on
the harder setting, none of
the classical methods tested reaches the optimum at the sampled sizes
from $n = 72$ onward within the tested budgets. 
While this does not provide evidence for quantum advantage, it identifies a domain
for further investigation.}
\keywords{QUBO, combinatorial optimisation, algorithm selection, tensor networks, QAOA, benchmarking}
\maketitle

\section{Introduction}
\label{sec:intro}

Quadratic unconstrained binary optimization (QUBO) provides a common
formulation for comparing classical, tensor-network, and quantum
approaches to combinatorial optimisation~\cite{Kochenberger2014,Lucas2014}.
For a practitioner, choosing a method requires specifying the desired
outcome: a low-energy solution, an optimum, or a certificate that no
better solution exists. These objectives need not favour the same
solver. A meaningful comparison must therefore distinguish solution
quality from optimality certification and relate both to the
computational budget.
These considerations connect to algorithm selection~\cite{Bischl2016},
which has been investigated for classical QUBO portfolios~\cite{Dunning2018}. 
Here, we focus on comparing solution quality, optimality certification, 
and computational cost across solver approaches, and on understanding 
how instance structure and trajectory diagnostics relate to their performance.

Problem instance structure offers a possible basis for making this choice.
Properties of the interaction graph can constrain the cost of exact
methods and may indicate the resources required by tensor-network
algorithms~\cite{MarkovShi2008,Schollwoeck2011}. Quantum-information
diagnostics offer a complementary perspective, but their interpretation
requires care. Alongside entanglement, we measure non-stabilizerness,
often called \emph{magic}: a quantum resource that characterises
departures from stabilizer states, which admit efficient classical
simulation~\cite{Gottesman1998,AaronsonGottesman2004,Leone2022}.
Magic concerns the resources needed for stabilizer-based simulation but
high magic alone (e.g., with low entanglement) 
does not rule out efficient classical simulation by 
methods that do not rely on a stabilizer representation.
Every QUBO optimum can be represented by a computational-basis state
with neither entanglement nor magic.
The relevant resources consequently arise in the intermediate states
explored by an algorithm, rather than in the optimal bitstring
itself~\cite{Dupont2022entanglement,Dupont2022calibrating,Nonstab2025qaoa}.
We study these resources along a transverse-field interpolation
connecting a simple initial Hamiltonian to the QUBO
Hamiltonian~\cite{KadowakiNishimori1998,AlbashLidar2018}.

This paper investigates how much structural and trajectory information
can explain solver performance, and whether it supports useful choices
between methods. We compare a general-purpose branch-and-cut solver
(Gurobi)~\cite{Gurobi}, simulated annealing
(SA)~\cite{Kirkpatrick1983}, a density-matrix renormalization group
(DMRG) based tensor-network (TN) anneal~\cite{White1992,Schollwoeck2011},
and fixed-schedule linear-ramp quantum approximate optimization
(LR-QAOA)~\cite{Farhi2014,MontanezBarrera2025}, with parallel tempering
(PT)~\cite{SwendsenWang1986,HukushimaNemoto1996} providing an additional classical comparison. Each serves a
specific role: branch-and-cut supplies a baseline capable of
certification, SA supplies a simple thermal-search baseline and PT
tests whether SA's failures persist under a stronger thermal method.
The DMRG anneal probes variational optimisation of matrix-product
states (MPS), while LR-QAOA probes a quantum-circuit approach with prescribed
parameters. The shared Hamiltonian interpolation connects these last
two methods, although their computational procedures and intermediate
states remain distinct.

Our main benchmark (Campaign A) comprises 180 instances from ten problem
families with $n = 14$--$48$ variables. It includes 90 small calibration
instances with exhaustively verified optima and 90 larger instances used to
assess extrapolation; the latter also receive an exact-solver
cross-check (Sec.~\ref{sec:groundtruth}). The robustness controls
(Campaign B) and scaling study (Campaign E) also use the
\emph{Wishart-planted} family~\cite{Hamze2020wishart}: synthetic problems
constructed from random Gaussian vectors so that a chosen solution is
guaranteed to be optimal, with a parameter controlling problem difficulty.
These known optima provide reference targets beyond the range of exhaustive
enumeration. To avoid giving solvers an easily guessed target, we randomly
complement selected binary variables ($x_i \mapsto 1-x_i$) and transform
the objective accordingly. This relabelling, called \emph{gauge
randomisation}, moves the planted optimum from the all-zero bitstring to
a random bitstring while preserving the energy landscape. It controls
for the advantage a solver may gain from the original placement of the
solution~\cite{MandraKatzgraber2018}. These experiments examine the
robustness of the comparison and extend the study to $n = 128$
(Secs.~\ref{sec:controls} and~\ref{sec:frontier}).

The results distinguish three aspects of solver performance.
First, branch-and-cut provides a strong baseline in Campaign A, matching every reference energy and certifying nearly all
instances. Its advantage nevertheless depends on whether the task is
search or certification: on the harder gauge-randomised instances, PT
can find optima that branch-and-cut misses within its allotted budget
(Secs.~\ref{sec:controls} and~\ref{sec:frontier}).

Second, a cheap treewidth bound of the QUBO interaction graph ranks
the bond-dimension demand of the tensor-network trajectories, but
supplies only a coarse runtime estimate. The evaluated structural
features and cost models do not produce a useful solver-selection
rule (Sec.~\ref{sec:advisor}). Controlled experiments further identify
excited-state trapping during warm-started DMRG sweeps as a failure
mechanism that can be mitigated by restarting the variational search
(Sec.~\ref{sec:regionT}). Entanglement and magic are only moderately correlated
in the measured trajectories and provide distinct diagnostic
information.

Third, the tested LR-QAOA runs provide no solution-quality improvement
over the classical comparators (Sec.~\ref{sec:regionQ}). The larger
Wishart experiments identify instances on which the tested classical
methods struggle, offering a concrete setting for further
quantum--classical comparisons. 
Together, these findings establish a classical baseline,
clarify what trajectory diagnostics can explain, and identify the
additional evidence needed for meaningful solver selection across
computational approaches.

Section~\ref{sec:methods} defines the shared framework and summarises
Campaigns A--E. Section~\ref{sec:taxonomy} presents the main benchmark
and robustness controls (A and B). Section~\ref{sec:regionT} brings
together the benchmark diagnostics and targeted TN mechanism tests (C).
Section~\ref{sec:regionQ} reports the LR-QAOA comparison and supplementary
quantum studies (D), followed by the Wishart scaling study (E) in
Sec.~\ref{sec:frontier}. Supporting experiments and exploratory resource
projections are collected in App.~\ref{app:supporting}.

\section{Problem formulation and experimental framework}
\label{sec:methods}

This section defines the optimisation problem, the benchmark instances,
and the common evaluation framework. We distinguish solution quality,
optimality certification, and computational cost. The main benchmark
provides the basis for the solver comparison and predictive analyses;
additional campaigns examine robustness, tensor-network failure
mechanisms, LR-QAOA protocols, and scaling to harder instances.
Section~\ref{sec:campaigns} summarises their questions and coverage.

\subsection{QUBO formulation and Hamiltonian interpolation}
\label{sec:setup}

We consider QUBO instances defined by a real symmetric matrix
$Q \in \mathbb{R}^{n \times n}$, with the objective
\[
  E^{*} = \min_{x \in \{0,1\}^n} x^T Q x.
\]
Introducing classical spin variables $z_i \in \{-1,1\}$ through
$x_i = (1-z_i)/2$ gives the Ising form
\begin{equation}
  E(z) = \sum_{i<j} J_{ij} z_i z_j + \sum_i h_i z_i + c,
  \label{eq:ising}
\end{equation}
where $J_{ij} = Q_{ij}/2$,
$h_i = -Q_{ii}/2 - \tfrac12 \sum_{j \ne i} Q_{ij}$, and
$c = \tfrac12\sum_i Q_{ii} + \tfrac12\sum_{i<j} Q_{ij}$
for this symmetric-matrix convention~\cite{Lucas2014}.
The additive constant preserves the objective value but does not
affect which bitstrings minimise it.

Replacing each spin variable by the Pauli-$Z_i$ operator gives the
diagonal cost Hamiltonian
\[
  H_Z = \sum_{i<j} J_{ij} Z_i Z_j + \sum_i h_i Z_i + c I,
\]
where $I$ is the identity operator. We retain the constant here so
that the eigenvalue associated with $\lvert x\rangle$ equals $x^TQx$.
Every optimal bitstring defines a computational-basis ground state of
$H_Z$, with zero entanglement and zero magic. A degenerate ground
space may also contain superpositions with nonzero quantum resources.
Thus the resources of an optimal bitstring alone do not distinguish
instance difficulty; we instead examine the intermediate states
produced by specified computational procedures~\cite{Dupont2022entanglement,
Dupont2022calibrating,Nonstab2025qaoa}.

The Hamiltonian interpolation used in this study is
\begin{equation}
  H(s) = -(1-s) \sum_i X_i + s\, H_Z, \qquad s \in [0, 1],
  \label{eq:anneal}
\end{equation}
where $X_i$ is the Pauli-$X$ operator on spin $i$.
At $s=0$, the ground state is the product state
$\lvert+\rangle^{\otimes n}$, with
$\lvert+\rangle=(\lvert0\rangle+\lvert1\rangle)/\sqrt{2}$;
at $s=1$, the Hamiltonian is $H_Z$.

This interpolation provides a common Hamiltonian framework for the
methods considered here. Quantum annealing evolves under a
time-dependent Hamiltonian along this path. LR-QAOA uses alternating
mixer and cost unitaries related to a discretisation of that evolution,
with the coefficient normalisation and schedule specified in
App.~\ref{app:solvers}. The DMRG anneal instead performs successive
variational optimisations at selected values of $s$. These procedures
need not produce the same intermediate states. In particular, the
entanglement measured along a DMRG sweep does not by itself bound the
entanglement of the corresponding LR-QAOA circuit.

The interpolation thus connects the computational approaches without
identifying their intermediate states. We next specify the instances
on which they are compared and the reference energies used to evaluate
their outputs.

\subsection{Instance families and benchmark selection}
\label{sec:families}

The ten families in Table~\ref{tab:families} cover graph sparsity,
frustration, penalty encodings, ground-state degeneracy, and planted
solutions. Sherrington--Kirkpatrick (SK) instances provide a dense
random spin-glass ensemble. Wishart-planted
instances~\cite{Hamze2020wishart} have known optima by construction
and a difficulty parameter $\alpha=M/n$, where $M$ is the number of
Gaussian sample vectors. The main benchmark uses $M=\lfloor0.7n\rfloor$,
so $0.7$ is the nominal aspect ratio; Sec.~\ref{sec:frontier} varies
this parameter. The ferromagnetic chain
provides a control with treewidth one and an analytic optimum.
The Max-Cut random graphs are unweighted. Maximum independent set (MIS) uses edge probability
$0.3$ and penalty weight $2$. Exact cover uses $n$ sets over a universe
of $n-4$ elements, including a hidden partition. SK couplings have
standard deviation $1/\sqrt{n}$ and zero fields. The random-frustrated
generator uses coupling density $0.8$, equally likely signs with
half-normal magnitudes, and Gaussian fields of standard deviation $0.1$.
The ferromagnetic chain has unit coupling strength.

\begin{table}[tbp]
\caption{Benchmark families and the regime each is included to probe.
The first column is the short handle used for the family throughout this
paper, including the benchmark and control comparisons;
MIS = maximum independent set, ER = Erd\H{o}s--R\'enyi random graph with
edge probability $p_{\mathrm{ER}}$ (written with a subscript because $p$
denotes the QAOA depth throughout, Sec.~\ref{sec:solverdesc}).
Requested generator sizes are $\{14,18,22\}$ (calibration) and
$\{32,40,48\}$ (out of sample), with seed indices $0,1,2$.
Actual sizes differ only for the grid family in this benchmark
(Sec.~\ref{sec:split}).}
\label{tab:families}
\begin{tabular}{lll}
\toprule
family & structure & probes \\
\midrule
ferro chain & chain, $\mathrm{tw}=1$, unfrustrated & control; analytic optimum \\
grid & Max-Cut, planar, bipartite & low treewidth; analytic optimum \\
3-regular & Max-Cut, sparse, degree 3 & high ground-state degeneracy \\
sparse ER & Max-Cut, $p_{\mathrm{ER}} = 0.2$ & moderate treewidth \\
dense ER & Max-Cut, $p_{\mathrm{ER}} = 0.5$ & treewidth $\sim n$ \\
MIS & ER, dense penalties & penalty-encoding overhead \\
exact cover & dense penalties & penalty encoding, planted solution \\
SK & dense random spin glass & random frustrated couplings \\
random frustrated & dense, tunable signs & frustration, no planted optimum \\
Wishart & dense, broad coupling magnitudes & planted optimum, tunable hardness \\
\bottomrule
\end{tabular}
\end{table}

\subsubsection{Calibration and out-of-sample evaluation}
\label{sec:split}

The main benchmark comprises 180 benchmark instances, indexed by problem
family, requested size, and instance-generation seed. This count follows
the benchmark's family--size--seed combinations; it does not imply 180
distinct problem matrices. We use \emph{distinct instances} when matrix
uniqueness matters and \emph{runs} for solver executions. The benchmark
is split into two sets by instance size:
\begin{itemize}
\item The \emph{calibration set} contains 90 instances at requested
sizes $n_{\mathrm{req}}\in\{14,18,22\}$. The grid family has actual
sizes $n\in\{15,20,24\}$; all other families retain the requested
sizes, giving an overall range $n=14$--$24$.
All reference optima are verified by
exhaustive enumeration.
This set is used to fit the runtime models
in App.~\ref{app:costmodels}.
\item The \emph{out-of-sample set} contains 90 instances at requested
sizes $n_{\mathrm{req}}\in\{32,40,48\}$. The grid family has actual
sizes $n\in\{35,42,48\}$; all other families retain the requested sizes.
Their reference energies are known by
construction or analytically, or certified by the exact-solver
cross-check, except for one SK instance whose reference remains
best-known (Sec.~\ref{sec:groundtruth}).
\end{itemize}

For a requested size $n_{\mathrm{req}}$, the grid generator uses
$a=\lfloor\sqrt{n_{\mathrm{req}}}\rfloor$ rows and
$b=\lceil n_{\mathrm{req}}/a\rceil$ columns, giving $n=ab$ variables.
This is a generator convention, not a restriction on permissible grids.
The 3-regular generator increases odd requested sizes by one, but
all requested sizes in this benchmark are even.

The grid and ferromagnetic-chain families have no random parameters, so
their generators do not depend on the seed. For these families, the
three benchmark instances at each size share the same matrix. Solving
them therefore repeats the computation on that matrix rather than
sampling three independently generated problems. Each split thus
contains 78 distinct matrices, 156 in total. Rank correlations between
instance features are reported over the distinct matrices, since the
repeated matrices would otherwise enter them three times.

This split tests for size extrapolation rather than interpolation within
a randomly partitioned size range. The reported out-of-sample runtime
errors use models fitted on the calibration set. Exploratory analyses
using all instances, including success-probability weighting, are
identified separately in App.~\ref{app:costmodels}.

\subsubsection{Wishart extensions and optimum placement}

The additional Wishart studies use known planted optima to evaluate
search beyond the range of exhaustive enumeration. The aspect ratio
$\alpha=M/n$ controls the ensemble, with the main benchmark using
$M=\lfloor0.7n\rfloor$ and the scaling study also considering
$\alpha=0.5$ (Sec.~\ref{sec:frontier}).

The original construction of these instances places an optimum at the all-zero bitstring.
To test sensitivity to this special placement, we use gauge-transformed
variants: selected binary variables are complemented and the objective
and reference energy are transformed accordingly. This moves the planted
optimum to a generic bitstring while preserving the landscape up to
relabelling and an accounted-for additive energy shift. Original and
gauged variants form a paired comparison, not independent instance
samples. The control subsets and repetitions are specified in
Sec.~\ref{sec:controls}; the larger scaling study has its own instance
set and budgets.

\subsection{Reference energies and optimality status}
\label{sec:groundtruth}

We denote $E^{*}=\min_{x\in\{0,1\}^n}x^TQx$ as the true global
minimum and additionally denote the benchmark target by $E_{\mathrm{ref}}$.
This reference energy is the objective value against which a solver's
returned solution is scored. When it is established by exhaustive
enumeration, solver certification, an analytic argument, or a planted
construction, $E_{\mathrm{ref}}=E^{*}$. Otherwise it is the best-known
feasible value, so $E_{\mathrm{ref}}\ge E^{*}$. 
A hit of a solver means matching $E_{\mathrm{ref}}$ within the
tolerance defined in Sec.~\ref{sec:metrics}; it implies reaching the
optimum only when the reference is known to be optimal.
Solver-certified references use the numerical optimality convention
specified in Sec.~\ref{sec:solverdesc}.
The reference construction and its optimality status depend on size
and family.

For all instance families at $n \le 24$, exhaustive enumeration returns the exact
optimum with a proof. All 90 calibration instances are of this kind.

For Wishart, grid, exact cover, and ferro-chain instances at all sizes,
the optimum is planted or known analytically, independently of any solver:
the Wishart ensemble is generated around a chosen ground-state
configuration~\cite{Hamze2020wishart}; a bipartite
grid Max-Cut has $E^{*} = -m$, where $m$ is the number of edges
(every edge can be cut); the exact-cover instances
are built around a hidden exact partition; and the ferromagnetic chain is
minimised by the aligned configuration. These constructions provide
$E_{\mathrm{ref}}=E^{*}$ for 36 of the 90 out-of-sample instances,
without requiring exhaustive enumeration.

Of the remaining 54 out-of-sample instances (dense and sparse ER,
3-regular, MIS, random frustrated, and SK), 53 have reference energies
certified as optimal by branch-and-cut. The sole exception is one SK
instance at $n=48$. For this instance, $E_{\mathrm{ref}}$ is the
best-known feasible energy, reached by both an extended SA run
and branch-and-cut, but not certified within the exact solver's time
limit. The extended SA budget is specified in Sec.~\ref{sec:solverdesc}.

Thus $E_{\mathrm{ref}}=E^{*}$ is established for 179 of the 180 main
benchmark instances: 90 by enumeration, 36 larger instances by
construction or analytic arguments, and the remaining 53 by the
exact-solver check. Equality remains unverified only for the SK
instance identified above. In the additional Wishart experiments,
including the larger scaling study, the references remain known
optima by construction, independently of solver success.

\subsection{Solver approaches and main-benchmark configurations}
\label{sec:solverdesc}

We compare an exact optimisation baseline with thermal-search and tensor-network heuristics, 
and a quantum-circuit heuristic. Table~\ref{tab:budgets}
summarises their computational budgets. The descriptions below explain
the choices needed to interpret the comparison; numerical settings and
implementation checks are collected in App.~\ref{app:solvers}.

\paragraph{Exact solver.}
Branch-and-cut searches restricted subproblems and uses bounds,
strengthened by additional constraints, to discard regions that cannot
improve the best solution found and ultimately certify optimality.
We use Gurobi~\cite{Kochenberger2014,Gurobi}, which minimises the binary quadratic
objective, using either a linearised or a native quadratic formulation,
with a $600$\,s time limit. Finding the reference energy and certifying
optimality are distinct outcomes: the reported certificate flag denotes
the solver's optimal status under its numerical tolerances
(Sec.~\ref{sec:groundtruth}). Exhaustive enumeration provides an
independent cross-check at small sizes. This general-purpose baseline
does not establish the limits of exact optimisation: specialised
branch-and-cut~\cite{Rehfeldt2023} or methods exploiting low treewidth
may perform better.

\paragraph{Simulated annealing.}
SA~\cite{Kirkpatrick1983} explores bitstrings through single-variable
flips, accepting some energy increases to escape local minima.
Metropolis acceptance becomes less permissive as the temperature falls.
We use a geometric cooling schedule. 
Each run comprises four restarts of 2000 sweeps,
where a sweep proposes one update per variable. The best state from
each restart is refined by single-flip local search, and the lowest
energy is returned. The extended runs used to establish reference
energies receive four times this update budget.

\paragraph{Parallel tempering.}
Similar to SA, PT~\cite{SwendsenWang1986,HukushimaNemoto1996} runs several searches at different temperatures and exchanges
configurations between them, allowing low-temperature searches to
benefit from exploration at higher temperatures. As a second
thermal-search baseline, we use 16 replicas with exchange attempts
after every sweep.
It uses a wall-clock budget and receives no final local-search polish.
It is reported alongside SA but is not included in the retrospective
heuristic winner labels or assigned a fitted runtime model.
The main and control budgets are summarised in Table~\ref{tab:budgets}
and Sec.~\ref{sec:controls}.

\paragraph{Tensor-network anneal.}
The TN anneal approximately minimises the interpolating Hamiltonian
within a compact MPS representation, using each state as the starting
point for the next interpolation step. We use two-site
DMRG~\cite{White1992,Schollwoeck2011}, with bond dimension capped
at $\chi=32$. The bond dimension controls the
size of the MPS representation and limits the entanglement it can
represent. Cold-start checks mitigate
excited-state trapping (Sec.~\ref{sec:regionT}). Several candidate
bitstrings are extracted from the final MPS, each is refined by
single-flip local search, and the lowest energy is returned.
Separate runs at $\chi=4,8,16$ assess sensitivity to the bond cap.
These optimisation runs all enable reset checks.
The trajectory diagnostics use separate, reset-free sweeps (Sec.~\ref{sec:tier1}).

\paragraph{LR-QAOA.}
LR-QAOA prepares a quantum state by alternating operations determined
by the objective with operations that mix computational-basis states,
then samples bitstrings as candidate solutions. It uses fixed linear
parameter schedules on a
coefficient-normalised Hamiltonian~\cite{MontanezBarrera2025}.
Here, a \emph{quantum circuit} is the sequence of quantum gates that
prepares the state before sampling. All LR-QAOA results reported in
this paper are obtained by classical simulation of such circuits,
none are quantum-hardware executions.
The number of LR-QAOA layers is denoted by $p$, and $\Delta$ sets
the amplitude of the linear parameter schedules. With $m$ nonzero
pairwise couplings, the LR-QAOA circuit uses $pm$ logical two-qubit cost gates.
The layer count $p$ differs from the gate-level circuit depth $D$,
which also accounts for how gates within a layer are scheduled
(App.~\ref{app:quantumexploration}).
Campaign A uses exact statevector simulation for $n\le16$
and approximate MPS simulation for $16<n\le32$, subject to a limit
of 3000 logical two-qubit cost gates. Table~\ref{tab:budgets} gives
the depths and sampling budgets. Runs replaced by SA because of
these limits are excluded from quantum comparisons. The lowest-energy
sampled bitstring is selected and then polished once by single-flip
local search. 
Unlike TN, this does not polish every candidate.
The separate depth-scaling and adaptive-depth studies are discussed
in Sec.~\ref{sec:regionQ}.

Agreement between the statevector and MPS simulators at small size checks
implementation conventions, but does not establish convergence of
the larger MPS simulations. Nor do the DMRG trajectory diagnostics
bound entanglement in the LR-QAOA states (Sec.~\ref{sec:setup}). The MPS
results must therefore be interpreted as approximate simulations of LR-QAOA.

\paragraph{Timings and interpretation.}
All timings were measured on the same machine.
Details about hardware and software are documented in App.~\ref{app:repro}.
Wall-clock comparisons reflect both algorithmic performance and 
implementation choices, including threading and compiled numerical routines. 
LR-QAOA simulation times measure classical computational cost, 
not quantum hardware runtime.
Circuit depth is reported separately 
as a resource estimate rather than a complete time-to-solution measure.

\begin{table}[tbp]
\caption{Budgets and applicability for the main benchmark comparison.
Budgets specify the work or elapsed time allocated to one run.
Applicability specifies which instances are eligible. Repetitions are
runs per eligible instance in this comparison, excluding the separate
SA repeat control and TN repeat check (Sec.~\ref{sec:controlresults}). PT uses a fixed $0.75$\,s budget per run,
not a budget matched separately to each SA run. The LR-QAOA rows describe Campaign A while supplementary LR-QAOA studies
are described in Sec.~\ref{sec:regionQ} and App.~\ref{app:circuitsupport}.
Here $\chi$ is the MPS bond-dimension cap, $p$ the number of LR-QAOA
layers, $\Delta$ the linear-schedule amplitude, and $m$ the number
of nonzero pairwise couplings. Implementation settings are given
in App.~\ref{app:solvers}.}
\label{tab:budgets}
\footnotesize
\begin{tabularx}{\textwidth}{@{}YYYY@{}}
solver & budget per run & applicability & repetitions \\
\midrule
exact branch-and-cut & up to $600$\,s; stops earlier at solver optimal status & all instances & $1$ \\
simulated annealing & $4$ restarts $\times$ $2000$ sweeps ($8000n$ proposals) & all instances & $1$ \\
parallel tempering & fixed $0.75$\,s; $16$ replicas & all instances & $10$ \\
DMRG anneal (TN) & $\chi=32$; $11$ interpolation points, ending at $s=1$ & all instances & $1$ \\
LR-QAOA, statevector & $p\in\{16,64\}$, $\Delta\in\{0.3,0.5,0.7\}$; $3996$ samples total ($666$ per setting) & $n\le16$ & $1$ \\
LR-QAOA, MPS proxy & $p=12$, $\Delta=0.5$, $\chi=32$; $500$ samples & $16<n\le32$ and $pm\le3000$ logical two-qubit cost gates & $1$ \\
\bottomrule
\end{tabularx}
\end{table}

\subsection{Diagnostic quantities and evaluation measures}
\label{sec:diagnostics}

The diagnostics are organised in three \emph{tiers} ---
\emph{structural}, \emph{trajectory}, and \emph{spectral} --- ordered by
what they cost and therefore by how far up in $n$ they remain available.

\begin{itemize}
\item \emph{Tier 0} is read off $Q$ and its interaction graph alone, costs
$O(m)$--$O(n^3)$ and does not require a quantum-state representation
(Sec.~\ref{sec:tier0}). This tier is intended for inexpensive pre-solve
assessment; the question of what the expensive tiers buy recurs in
Sec.~\ref{sec:advisor}.
\item \emph{Tier 1} requires a matrix-product-state
sweep along the trajectory $H(s)$ of Eq.~\eqref{eq:anneal} and is
affordable to $n \lesssim 10^2$ (Sec.~\ref{sec:tier1}). 
\item \emph{Tier 2} is exact diagonalisation, limited to $n \lesssim 20$ (Sec.~\ref{sec:tier2}).
It is used to validate and interpret the other diagnostics, rather than
as an input to the cost models or the exploratory indicator of
instances for further quantum testing (Eq.~\eqref{eq:flag},
App.~\ref{app:quantumexploration}). 
\end{itemize}

Table~\ref{tab:measures} collects the measured quantities, both the
instance diagnostics and the outcome metrics, in one place: what each
one captures, what it costs, and where in the paper it is used, with definitions given below.

\begin{table}[tbp]
\caption{The measures used in this paper: instance
diagnostics (Tiers~0--2), obtained from the instance and the specified
diagnostic computations, and outcome metrics recorded per solver run. Costs are per
instance, $m$ is the number of couplings, $n_s$ the number of grid points
in $s$, and $\chi$ the matrix-product-state bond dimension. The reference
energy $E_{\mathrm{ref}}$ against which all hit-based metrics are scored is
certified, planted, or best-known depending on size
(Sec.~\ref{sec:groundtruth}).}
\label{tab:measures}
\footnotesize
\begin{tabularx}{\textwidth}{@{}lYYY@{}}
measure & what it captures & how obtained (cost) & where it is used \\
\midrule
\multicolumn{4}{l}{\emph{Tier 0 --- structure and landscape of $Q$; any $n$, cost $O(m)$--$O(n^3)$ (Sec.~\ref{sec:tier0})}} \\[2pt]
$\widehat{\mathrm{tw}}$ & distance from a tree; bounds the exponent of exact TN contraction & min-degree/min-fill elimination bound ($O(n^3)$, the one costly Tier-0 feature) & TN cost law, Eq.~\eqref{eq:ttn}; exactness rule R3 (App.~\ref{app:costmodels}) \\
$\rho_G$ & edge density $2m/[n(n-1)]$ & from the interaction graph & quantum indicator, Eq.~\eqref{eq:flag} \\
$b_{\mathrm{RCM}}$ & distance from a one-dimensional chain & bandwidth after reverse Cuthill--McKee (RCM) reordering~\cite{CuthillMcKee1969} & secondary correlate of $\chimax$ (Sec.~\ref{sec:regionT}) \\
$D_J$ & coexisting energy scales, Eq.~\eqref{eq:dj} & $\max |J_{ij}| / \min |J_{ij}|$ & quantum indicator \\
balance, planarity, $f_\triangle$ & three cheap graph predicates: frustration exact and graded, and polynomial exactness & parity 2-colouring~\cite{Harary1953}; planarity test~\cite{Hadlock1975}; sampled triangles~\cite{Toulouse1977} & exactness rules R2, R4 (App.~\ref{app:costmodels}) \\
$h_{\mathrm{LM}}$ & ruggedness: multi-start local-minimum hit rate & fraction of $50$ first-improvement descents reaching their best value & quantum indicator \\[2pt]
\midrule
\multicolumn{4}{l}{\emph{Tier 1 --- barriers along the trajectory $H(s)$, Eq.~\eqref{eq:anneal}; $n \lesssim 10^2$ (Sec.~\ref{sec:tier1})}} \\[2pt]
$\Smax$ & entanglement barrier: peak von Neumann entropy over all cuts and all $s$ & warm-started DMRG sweep at $\chi = 16$ ($O(n_s n \chi^3)$) & classical-hardness axis; Page scale $n/2$~\cite{Page1993} \\
$\chimax$ & maximum number of Schmidt components retaining $99\%$ weight & same sweep & $\widehat{\mathrm{tw}} \to \hat\chi$ law (Sec.~\ref{sec:regionT}); Eq.~\eqref{eq:ttn} \\
$\Mmax$ & magic barrier: peak non-stabilizerness, Eq.~\eqref{eq:m2} & four-replica MPS contraction~\cite{HaugPiroli2023} at $\chi = 4$ ($O(n\chi^{10})$); truncated estimate & complementary hardness indicator; quantum indicator \\[2pt]
\midrule
\multicolumn{4}{l}{\emph{Tier 2 --- exact spectra; $n \lesssim 20$; validation and interpretation (Sec.~\ref{sec:tier2})}} \\[2pt]
$\Delta_g(s)$ & $E_g(s)-E_0(s)$, using the final ground-state degeneracy $g$ & sparse diagonalisation of $H(s)$ over all $2^n$ states & stack validation; trapping analysis (Sec.~\ref{sec:regionT}); TN readout design (App.~\ref{app:solvers}) \\[2pt]
\midrule
\multicolumn{4}{l}{\emph{Outcome metrics --- measured per solver run (Sec.~\ref{sec:metrics})}} \\[2pt]
hit & reached the reference energy $E_{\mathrm{ref}}$ & Eq.~\eqref{eq:hit}, $\epsilon = 10^{-9}$ (mixed absolute/relative) & main benchmark and repeat controls; scaling exceptions below \\
proved & optimality \emph{certificate}, subject to numerical tolerances & solver optimal status within the stated cap & Secs.~\ref{sec:regionC}, \ref{sec:frontier} \\
$t$ & wall-clock seconds to the returned solution & per run; circuit depth reported separately & cost-model fits, Eqs.~\eqref{eq:tsa}--\eqref{eq:ttn} \\
$w^{*}$ & ex-post best solver, the choice made knowing all outcomes & lowest $E$, ties broken by smaller $t$, Eq.~\eqref{eq:expost} & retrospective heuristic-label analysis \\
$d_H^{\min}/n$ & how far a run ends from the nearest optimum & Hamming distance to the enumerated optimal set & trapping diagnostic (Sec.~\ref{sec:regionT}) \\
\bottomrule
\end{tabularx}
\end{table}

\subsubsection{\texorpdfstring{Structural features (Tier 0; cost $O(m)$--$O(n^3)$, any $n$)}{Structural features (Tier 0)}}
\label{sec:tier0}

These features use the instance data and a small local-search probe,
without a quantum-state simulation.

The interaction graph $G=(V,E_G)$ has one vertex per binary variable
and an edge $(i,j)$ wherever $J_{ij}\ne0$. Its edge count is $m=|E_G|$.

\emph{Treewidth bound $\widehat{\mathrm{tw}}$.} The treewidth of the
interaction graph measures how far the graph is from a tree: a tree
decomposition of width $w$ groups the vertices into overlapping ``bags''
of size $w+1$ arranged in a tree, such that every edge lies inside some
bag and the bags containing any given vertex form a connected
subtree. Its operational meaning here is that both exact dynamic programming
over such a decomposition and exact tensor-network contraction cost
$\exp[O(\mathrm{tw})]$~\cite{MarkovShi2008,Dumitrescu2018}: the
treewidth is the size of the ``interface'' a contraction must carry
along, and hence directly the exponent of the tensor-network cost.
Computing $\mathrm{tw}$ exactly is NP-hard, so we use the standard
min-degree/min-fill elimination heuristics and take the smaller bound,
$\widehat{\mathrm{tw}} \ge \mathrm{tw}$. The two are kept apart
throughout: $\mathrm{tw}$ is the exact treewidth, in which all cost
statements and the structural descriptions of the instance families are
phrased, while $\widehat{\mathrm{tw}}$ is the computed bound, the
quantity the pipeline actually reads. A chain has $\mathrm{tw} = 1$ and a
fully connected graph on $n$ vertices has $\mathrm{tw} = n-1$.

Two distinct tensor-network computations appear in this paper and should
not be conflated. The first is the \emph{exact} contraction just
described, equivalently dynamic programming over a tree decomposition,
which returns the optimum together with a proof of optimality at a cost
$\exp[O(\mathrm{tw})]$. It is what gives
$\widehat{\mathrm{tw}}$ its operational meaning, and it is the method
behind the bounded-treewidth rule R3 of App.~\ref{app:costmodels}:
once $\widehat{\mathrm{tw}}$ is small enough for this contraction to be
affordable, the instance can be settled exactly and no heuristic cost has
to be estimated.
The second is our \emph{solver}
(Sec.~\ref{sec:solverdesc}): a variational matrix-product-state anneal,
a heuristic with no optimality guarantee, whose cost is
\emph{polynomial} in a bond dimension $\chi$ that is not fixed by
$\mathrm{tw}$ a priori. That larger $\widehat{\mathrm{tw}}$ nevertheless
means larger realised $\chi$ is a measured result of this work
(Sec.~\ref{sec:regionT}), not a definition, and it is what licenses
putting $\widehat{\mathrm{tw}}$ into the solver's runtime model,
Eq.~\eqref{eq:ttn}.

The coupling dynamic range $D_J$ is the ratio of the strongest to the
weakest coupling, using $J_{ij}$ from Eq.~\eqref{eq:ising}:
\begin{equation}
  D_J \;=\; \frac{\max_{(i,j) \in E_G} |J_{ij}|}
                 {\min_{(i,j) \in E_G} |J_{ij}|},
  \label{eq:dj}
\end{equation}
A large ratio indicates separated coupling scales, which can make
single-spin thermal exploration difficult: temperatures that resolve
weak couplings may suppress rearrangements involving strong couplings.
In Campaign A this does not separate SA successes from failures
(Sec.~\ref{sec:regionC}). The Wishart scaling study instead varies the construction parameter
$\alpha=M/n$ (Sec.~\ref{sec:frontier}).
The implementation includes only couplings with $|J_{ij}|>10^{-15}$
and sets the ratio to one if none remain.

Four cheap graph quantities and one landscape probe complete Tier~0.
Density is
$\rho_G = 2m/[n(n-1)]$. The graph bandwidth $b_{\mathrm{RCM}}$ is taken after a reverse
Cuthill--McKee reordering~\cite{CuthillMcKee1969}, which clusters the
adjacency matrix's nonzeros around the diagonal. 
The resulting bandwidth
measures how far the interaction graph is from a one-dimensional chain,
and hence how well a matrix-product state, intrinsically 1-D once a site
ordering is fixed, can cover the couplings. Frustration enters twice, as
exact signed-graph balance~\cite{Harary1953} and as the sampled fraction
of frustrated triangles $f_\triangle$~\cite{Toulouse1977}. A cycle is
frustrated when no spin assignment satisfies all its bonds. Planarity is
tested because it admits a polynomial exact algorithm for field-free
Max-Cut~\cite{Hadlock1975}. The landscape probe is the multi-start
local-minimum hit rate $h_{\mathrm{LM}}$: 50 descents from random
bitstrings use seeds $\mathrm{seed}+1000r$, $r=0,\ldots,49$.
The rate is the fraction reaching the best energy among these
50 descents, using the same mixed tolerance as the hit criterion in
Eq.~\eqref{eq:hit}.
The triangle probe checks all triangles if there are at most 20,000,
or randomly samples 20,000 without replacement. 
A frustrated cycle has negative product of $\operatorname{sign}(-J_{ij})$;
graph balance tests the corresponding bond constraints.

\subsubsection{\texorpdfstring{Trajectory diagnostics (Tier 1; $n \lesssim 10^2$)}{Trajectory diagnostics (Tier 1)}}
\label{sec:tier1}

At each interpolation point in Eq.~\eqref{eq:anneal}, two-site
DMRG~\cite{White1992,Schollwoeck2011} seeks a low-energy MPS
approximation, initialised from the preceding point. The resulting
sequence defines the diagnostic trajectory. It is not guaranteed
to follow the instantaneous ground state: warm starting can retain
an excited branch, as investigated in Sec.~\ref{sec:regionT}.
The measured resources refer to these variational states, with their
specified site ordering and bond-dimension caps.

The entanglement probe uses $\chi=16$ and the magic probe uses
$\chi=4$, in separate DMRG sweeps on the nine-point uniform grid
$s\in\{0,0.125,0.25,\ldots,1\}$. Neither probe enables the
branch-reset mechanism used by the optimisation solver. Magic is
computed on the states retained from the $\chi=4$ sweep, while
$\Smax$ and $\chimax$ come from the dedicated $\chi=16$ sweep.
The diagnostic and solver trajectories thus differ in grid and
reset configuration.

The entanglement barrier is measured by cutting the chain at bond $b$ and
Schmidt decomposing, which gives probabilities $p_1 \ge p_2 \ge \dots$.
The von Neumann
entropy $S_{\mathrm{vN}}(b;s) = -\sum_a p_a \log_2 p_a$ measures how much
quantum correlation crosses that cut, and
$\chi_{99}(b;s) = \min\{k : \sum_{a \le k} p_a \ge 0.99\}$ is the number
of Schmidt components needed to retain $99\%$ of the weight across
that cut of the computed state. We record maxima over the sampled
interpolation points and all cuts, $\Smax$ and $\chimax$.
The largest recorded $\chimax$ is 13 in each split. Being below
the bond-dimension cap does not alone establish convergence. The
reset-enabled $\chi=32$ solver trajectories (Sec.~\ref{sec:solverdesc})
record values up to 20 and exceed the diagnostic value on 64 of 90
calibration and 77 of 90 out-of-sample instances; the two agree in rank
at Spearman $\rho=0.69$ and $0.73$. The diagnostic $\chimax$ is therefore
a ranking proxy for the representation demand of the solver's
trajectory, not a measurement of it.
For comparison,
a Haar-random $n$-qubit pure state has half-cut entropy approximately
$n/2$ bits~\cite{Page1993}, with bond-dimension requirements growing
exponentially in $n$. The measured trajectories can be compared with
this reference scale without assuming that low entanglement ensures
successful optimisation.

Entanglement alone does not decide classical
simulability. \emph{Stabilizer} states, those preparable by Clifford
circuits, can be maximally entangled yet are simulable in polynomial
time by the Gottesman--Knill theorem~\cite{Gottesman1998,AaronsonGottesman2004}.
What obstructs such stabilizer-based simulation is
\emph{non-stabilizerness}, or ``magic'': the departure of a state from
the stabilizer class. We quantify it with the stabilizer 2-R\'enyi
entropy~\cite{Leone2022}
\begin{equation}
  M_2(\lvert\psi\rangle) = -\log_2\Bigl( 2^{-n}
  \sum_{P} \langle\psi\rvert P \lvert\psi\rangle^{4} \Bigr),
  \label{eq:m2}
\end{equation}
the sum running over all $4^n$ Pauli strings. $M_2 = 0$ exactly for
stabilizer states (their Pauli spectrum is flat on the stabilizer
group) and grows with magic. It is computable exactly on an MPS by a
four-replica transfer contraction~\cite{HaugPiroli2023}, which is what
makes it usable here. The contraction is exact for the supplied MPS but
the state itself is an approximation. At the $\chi=4$ probe cap,
$\Mmax$ is therefore a truncated estimate. It is used in rank
correlations and in the 8-bit threshold of
Eq.~\eqref{eq:flag}, not as a converged value for the exact state. That
threshold has no independent motivation: it was set on the Campaign A
data themselves (App.~\ref{app:quantumexploration}).

\begin{figure}[tbp]
  \includegraphics[width=\linewidth]{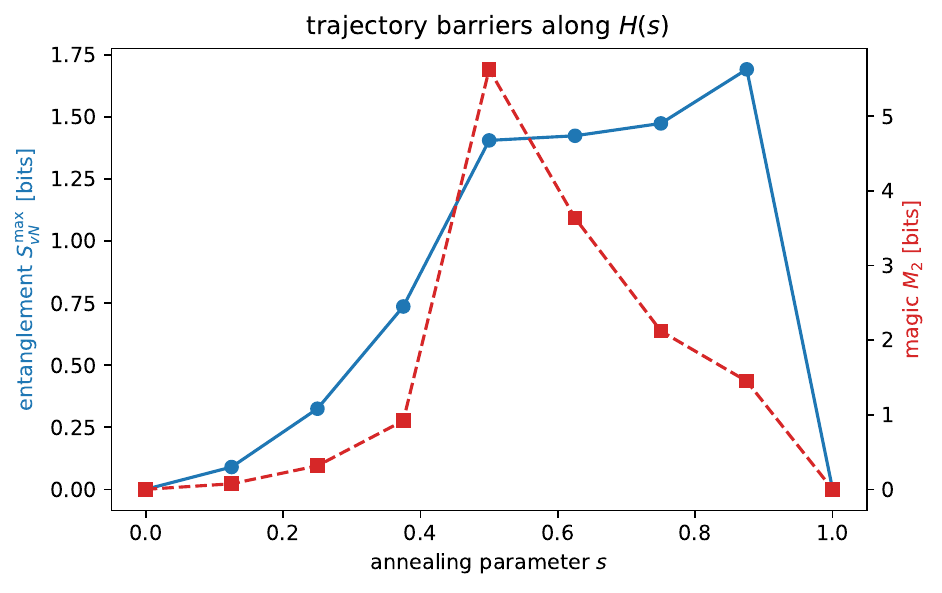}
  \caption{Trajectory diagnostics on an example instance
  (Erd\H{o}s--R\'enyi Max-Cut, $n = 16$, $m = 63$): entanglement barrier
  $S_{\mathrm{vN}}^{\max}(s)$ (left axis) and magic barrier $M_2(s)$
  (right axis) along $H(s)$, Eq.~\eqref{eq:anneal}, computed with the
  Tier-1 protocol ($\chi=16$ entanglement sweep, $\chi=4$ magic sweep,
  nine-point grid). Both vanish at the endpoints of this example. The two
  barriers peak at different points: $\Smax = 1.69$ bits at $s = 0.875$,
  $\Mmax = 5.63$ bits at $s = 0.5$ ($\chimax = 8$).}
  \label{fig:barriers}
\end{figure}

The initial state $\lvert+\rangle^{\otimes n}$ has zero magic.
A final computational-basis state also has zero magic, but this need
not hold for every superposition in a degenerate ground space.
Figure~\ref{fig:barriers} illustrates the measured profiles on one
instance. Magic and entanglement probe different aspects of state
representation. Neither a small measured magic value nor a large
one, on its own, establishes the cost of all classical simulation
methods.

\subsubsection{\texorpdfstring{Spectral diagnostics (Tier 2; $n \lesssim 20$)}{Spectral diagnostics (Tier 2)}}
\label{sec:tier2}

For small instances ($n\lesssim20$), sparse diagonalisation of
$H(s)$ provides an independent comparison with the DMRG energies
and resolves the levels used in the trapping analysis
(Fig.~\ref{fig:trapping}). This check applies only at the sizes
where the spectrum is computed.
The campaign-specific grids and eigensolver settings are given in
App.~\ref{app:tnsupport}.

Let $E_0(s)\le E_1(s)\le\cdots$ be the ordered eigenvalues, and let
$g$ denote the ground-state degeneracy of the final cost Hamiltonian
$H_Z$. At $s=1$, the first $g$ levels share the optimal energy, so
$E_1-E_0$ vanishes whenever $g>1$, independently of solver difficulty.
We record $\Delta_g(s)=E_g(s)-E_0(s)$ to avoid this endpoint
degeneracy in the reported diagnostic. At $s=1$ it is the gap above
the optimal manifold. At intermediate $s$, the first $g$ levels need
not be degenerate, so this definition should not be identified
automatically with the instantaneous ground-state excitation gap.

The $n \lesssim 20$ ceiling is the $2^n$ wall: the calculation lives in
the full Hilbert space ($10^6$ dimensions at $n = 20$), is repeated at
every point of the $s$-grid, and must resolve the entire $g$-fold
manifold \emph{plus} one level above it, so degeneracy multiplies the
eigenvalue count. Omitting it above $n = 20$ costs nothing operationally:
no Tier-0 feature, no cost model of Eqs.~\eqref{eq:tsa}--\eqref{eq:ttn},
and no term of the quantum indicator Eq.~\eqref{eq:flag} reads a spectral
quantity, so the cost predictions are identical whether or not spectra were
computed.

\subsubsection{Solver outcome and prediction metrics}
\label{sec:metrics}

For a solver returning energy $E$ in wall time $t$ we record whether it
reached the reference,
\begin{equation}
  \mathrm{hit} = \openone\!\left[\, |E - E_{\mathrm{ref}}| < \epsilon
  \max(1, |E_{\mathrm{ref}}|) \,\right], \qquad \epsilon = 10^{-9},
  \label{eq:hit}
\end{equation}
a mixed absolute/relative tolerance: for the typical instance with
$|E_{\mathrm{ref}}| \gg 1$ the criterion is relative ($\epsilon |E_{\mathrm{ref}}|$, so that
large energies are not penalised for floating-point noise), while the
$\max(1,\cdot)$ provides an absolute floor $\epsilon$ that keeps the test
well defined when $|E_{\mathrm{ref}}|$ is small or zero, where a purely relative
tolerance would demand unattainable precision. The indicator
$\openone(\cdot)$
is $1$ when its condition is true and $0$ otherwise.
Campaign E retains different reach tolerances across its run batches:
its exact-solver endpoint test and first-hit callback use a mixed
$10^{-6}$ tolerance, while its initial single-run SA test uses an
absolute $10^{-6}$ tolerance. Later SA repeats and PT use
Eq.~\eqref{eq:hit}. These conventions are retained when describing
stored results and first-hit times. Re-scoring all 67 stored Campaign E
SA records under each of the three tolerances changes no outcome: the
closest miss lies $0.43$ above the reference.
Beyond reaching $E_{\mathrm{ref}}$,
the exact solver additionally reports whether it proved optimality
(solver optimal status within the time cap, under the numerical
tolerances described in Sec.~\ref{sec:solverdesc}), a stronger outcome
than a hit and one no other solver here provides.

The per-instance ex-post best solver (within the evaluated heuristic portfolio), i.e., the choice one would have
made knowing all outcomes, is
\begin{equation}
  w^{*} = \arg\min_{P}\, E_P \quad
  \text{(ties broken by smaller $t_P$)},
  \label{eq:expost}
\end{equation}
with energy primary and time used only to break ties. This descriptive
label does not reward certification and is not a universal
cross-paradigm ranking. The retrospective heuristic analysis in
Sec.~\ref{sec:advisor} uses the Campaign A portfolio
$\{\mathrm{SA},\mathrm{TN},\mathrm{LR\mbox{-}QAOA}\}$ while PT is reported
separately. Any recorded LR-QAOA simulation time describes that
implementation, not quantum hardware performance. We therefore
report reference hits, certificates, and computational resources
separately rather than infer a hardware advantage from $w^*$.
The Campaign A label-generation code excludes failed runs and explicit
SA fallbacks in place of LR-QAOA runs. Energies less than $10^{-9}$ above the
lowest observed energy are treated as ties, broken by elapsed
solver-call time, including LR-QAOA simulation and polishing.
These labels compare software executions, not classical runtime
with quantum hardware runtime.

Association between a feature and an outcome is measured by the
tie-aware Spearman rank correlation $\rho$. For binary outcomes we use
the area under the receiver operating characteristic (ROC) curve,
abbreviated AUC. Let $\mathcal{P}$ and $\mathcal{N}$ denote
the positive and negative instances for the outcome being evaluated,
respectively, and let $u_i$ be the candidate predictor's score for
instance $i$, with larger scores indicating greater predicted likelihood
of the positive outcome.  
$|\mathcal{P}|$ and $|\mathcal{N}|$ are the numbers of instances in the
two sets. Then
\begin{equation}
  \mathrm{AUC} = \frac{1}{|\mathcal{P}||\mathcal{N}|}
  \sum_{i \in \mathcal{P}} \sum_{j \in \mathcal{N}}
  \Bigl[ \openone(u_i > u_j) + \tfrac12 \openone(u_i = u_j) \Bigr],
  \label{eq:auc}
\end{equation}
the probability that a randomly chosen positive instance scores above a
randomly chosen negative instance under the candidate predictor $u$:
$\mathrm{AUC} = 1$ is
perfect separation, $0.5$ is chance. We report it only when at least
three successes \emph{and} three failures are present. Finally, a
predicted cost $\hat t$ is scored against the measured time $t$ by the
symmetric fold error $\max(\hat t / t,\, t / \hat t)$
(Sec.~\ref{sec:advisor}).

We report correlations descriptively, without $p$-values. The
hypothesis thresholds and their outcomes are stated alongside the
corresponding analyses.

\subsection{Overview of the experimental campaigns}
\label{sec:campaigns}

The experiments are organised into five campaigns, summarised in Table~\ref{tab:campaigns}. 
Each campaign addresses a specific question through a defined choice of problem instances,
solver configurations, and repetitions. 
The overview distinguishes these experimental settings and identifies where campaigns share instances.

A problem instance is a QUBO matrix generated for a given family, size,
and seed. In Campaign A, the seed-independent grid and ferro-chain
families give the same matrix for all three seeds
(Sec.~\ref{sec:split}).
Within a campaign, a solver may be applied to the same instance several times.
Each execution with a specified configuration and, where applicable, random seed is called a \emph{run}. 

Some campaigns also compare an instance with a gauge-transformed variant, 
which relabels its variables while preserving the optimisation problem up to the corresponding energy shift. 
Repeated runs and gauge-transformed variants therefore do not constitute independently generated instances.

Campaign A provides the main benchmark and the calibration/evaluation split used for runtime modelling and solver-selection assessment. Campaigns B--E examine robustness, TN failure mechanisms, supplementary LR-QAOA protocols, and Wishart scaling, respectively.

\begin{table}[tbp]
\caption{Overview of the experimental campaigns. Sizes refer to actual
variable counts unless marked as requested. Shared instances and
restricted solver coverage are indicated explicitly. Main budgets are
in Table~\ref{tab:budgets}. Follow-up settings are given in the referenced
protocols and App.~\ref{app:solvers}.}
\label{tab:campaigns}
\footnotesize
\begin{tabularx}{\textwidth}{@{}lYYY@{}}
\toprule
 & question & instance set & solvers and computations \\
\midrule
A & Main comparison; diagnostic relationships and runtime prediction &
180 benchmark instances, ten families; requested sizes $14,18,22$ for
calibration and $32,40,48$ for evaluation &
Gurobi, SA, TN, and 105 simulated LR-QAOA runs; additional PT baseline;
structural and separate reset-free trajectory diagnostics
(Secs.~\ref{sec:families}--\ref{sec:metrics}) \\
\addlinespace
B & Sensitivity to stochastic variation and optimum placement &
SA repeats on all 180 benchmark instances; nine paired Wishart instances at
$n=32,40,48$; enlarged gauged set of ten per size at $n=40,48$ &
SA repetitions; Gurobi, SA, and PT controls; TN gauge rechecks on the
Campaign A $n=40,48$ subset and repeatability check on three ungauged
$n=40$ instances (Secs.~\ref{sec:controlresults} and~\ref{sec:controls}) \\
\addlinespace
C & TN capacity limits, trapping, and mitigation &
Main benchmark for bond-cap sensitivity; selected 3-regular and
ferromagnetic instances for mechanism studies &
TN bond-cap variants, spectral diagnostics, and comparisons of
multi-seeding and reset strategies
(Secs.~\ref{sec:tier2} and~\ref{sec:regionT}) \\
\addlinespace
D & LR-QAOA coverage, probability scaling, and simulation convergence &
27 benchmark instances at $n=32$; ten families at $n=8$--$16$;
six gauged Wishart instances at $n=40,48$ &
Adaptive-depth MPS LR-QAOA; exact statevector probability measurements;
bond-cap convergence probes using a different gauge convention from B
(Sec.~\ref{sec:regionQ}) \\
\addlinespace
E & Search and certification as hardness, size, and budget vary &
61 distinct gauged Wishart instances, $\alpha=0.5,0.7$, sampled sizes
$n=48$--$128$ &
Gurobi, SA, and PT at campaign-specific budgets; no TN or quantum solver
runs (Sec.~\ref{sec:frontier}) \\
\bottomrule
\end{tabularx}
\end{table}

The campaigns share some instances but evaluate different settings. 
Campaign B reuses Wishart instances from Campaign A for the paired 
gauge comparison and adds further instances for the enlarged comparison. 
Campaign D supplements the LR-QAOA results from Campaign A with tests 
of adaptive depth, simulation convergence, and success-probability scaling. 
Campaign E evaluates Gurobi, SA, and PT on a separate Wishart scaling set. 

\paragraph{Use of AI tools.}
AI assistance (Claude, Anthropic) was used for code development, running
analysis and verification scripts, and drafting; all results were checked
against tests, and the authors take full responsibility for the content.

\section{Main benchmark: solver performance and robustness}
\label{sec:taxonomy}

We first report the main benchmark (Campaign A), then examine
stochastic repetition and optimum-placement controls (Campaign B).
Original and gauged outcomes are reported separately because they use
different instance variants and replication schemes.

\subsection{Main benchmark comparison (Campaign A)}
\label{sec:regionC}

\begin{figure}[tbp]
  \includegraphics[width=\linewidth]{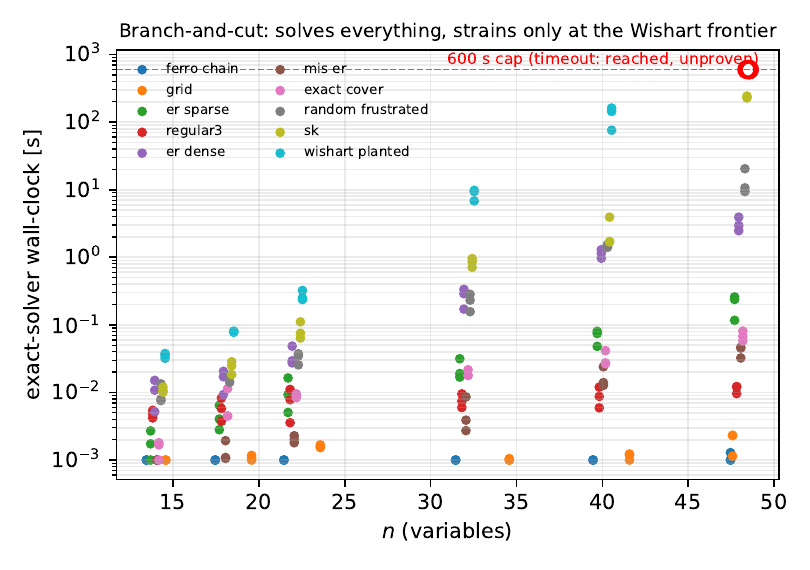}
  \caption{Exact branch-and-cut wall-clock vs.\ instance size, all $180$
  benchmark instances. Horizontal positions use actual variable counts with fixed
  family offsets between $-0.54$ and $0.54$ for visibility.
  Filled points have solver optimal status. Open red circles (all at
  $n = 48$, Wishart and one SK) matched the reference energy but hit
  the $600$\,s cap before certifying. The Wishart references are planted
  optima while the SK reference is best-known. Almost everything is solved in under $0.1$\,s, 
  only the dense frustrated families climb, and the Wishart instances at $n=48$
  strain certification. Campaigns B and E separately test search on
  gauged Wishart instances.
Timings reflect the solver implementation and execution environment; cross-solver runtime comparisons require the qualifications in Sec.~\ref{sec:threats}.}
  \label{fig:milp}
\end{figure}

The results for the exact solver are shown in Table~\ref{tab:outcomes} and Fig.~\ref{fig:milp}: the
exact branch-and-cut solver matches $E_{\mathrm{ref}}$ on all $180$
benchmark instances and certifies optimality on $176$, including $86$ instances at $n > 26$ that lie
beyond exhaustive enumeration, in a median of under a second (worst proven
case $\sim\!4$ minutes). Only the four densest frustrated instances at
$n = 48$ reach the $600$\,s cap while matching the reference energy.
Three of these references are known planted optima while the fourth, an SK
instance, only has a best-known solution. Simulated annealing also matches the
reference energies on every non-Wishart
instance, quickly but with no certificate, and misses six of the nine
out-of-sample Wishart instances. The DMRG tensor-network
anneal reaches the reference on all but two non-Wishart instances (one
sparse ER instance at $n=48$ and one 3-regular instance at $n=40$),
typically $10^2$--$3\times10^{3}\times$ slower than SA on the dense
families (median $940\times$), and misses seven of the
nine out-of-sample Wishart instances. The
LR-QAOA runs never improve on the better of SA and TN
(Sec.~\ref{sec:regionQ}). Campaign A establishes a strong certification baseline.
The repetition and optimum-placement controls below test how the search
comparison changes when stochastic variability and the special planted
bitstring are taken into account.

SA difficulty in Campaign A is confined to one family. On the
calibration set SA reaches every reference, so there is no variation in
difficulty for a Tier-0 feature to rank, and we report no
feature--difficulty correlation for it. Out of sample, all six SA misses
are Wishart instances. A wide coupling dynamic range $D_J$ alone does
not mark them: the SK and random-frustrated ensembles share it, and SA
solves every instance of both. On the out-of-sample Wishart instances,
SA does not consistently reach the planted optimum within its tested budget.
This motivates the repeated-run and optimum-placement controls 
in Campaign B (Sec.~\ref{sec:controls}).
Campaign E separately examines how performance changes with 
instance size and the Wishart aspect ratio \(\alpha=M/n\) (Sec.~\ref{sec:frontier}).

\begin{table}[tbp]
\caption{Main benchmark outcomes (Campaign A). Each of the nine
out-of-sample Wishart instances has one Campaign A SA run and one TN run. Their counts
are not repetition-based success estimates. Gurobi matches every
reference, but one SK reference remains best-known. LR-QAOA coverage
and family-specific misses are reported in Sec.~\ref{sec:regionQ}.}
\label{tab:outcomes}
\begin{tabularx}{\textwidth}{@{}lYY@{}}
\toprule
method & evaluation set & outcome \\
\midrule
Gurobi & all 180 & 180 reference hits; 176 solver certificates \\
SA & all 180 & hits on all 162 non-Wishart instances; out-of-sample Wishart: $2/3$ at $n=32$, $1/6$ at $n=40,48$ \\
TN & all 180 & hits on 160 of 162 non-Wishart instances (misses: sparse ER $n=48$, 3-regular $n=40$); out-of-sample Wishart: $1/3$ at $n=32$, $1/6$ at $n=40,48$ \\
LR-QAOA & 105 LR-QAOA simulation runs & $86/96$ non-Wishart hits; $5/9$ Wishart hits; no improvement over the better of SA and TN \\
\bottomrule
\end{tabularx}
\end{table}

\subsection{Sensitivity to stochastic repetition (Campaign B)}
\label{sec:controlresults}

Because SA is stochastic, a single run may not reliably characterise its 
performance on an instance. To assess this variability and estimate 
reference-hit rates, we repeat SA at the main budget using ten seeds 
on each of the 180 benchmark instances (1800 scored runs). Separate
runs at the Campaign A seeds reproduce all 180 recorded energies.
These checks are excluded from the repeat hit rates. 

SA reaches the optimum in all 900 benchmark runs on the 
calibration instances and in at least $98\%$ of runs on non-Wishart out-of-sample instances.

For Wishart, the repetitions qualify the Campaign A single-run outcomes. 
For the six ungauged Campaign A Wishart instances at $n=40,48$,
32 of 60 additional SA runs reach the planted optimum ($53.3\%$),
compared with one of six Campaign A runs ($16.7\%$).
This repeat control shows that the Campaign A single-run outcomes
understate SA's success rate on these instances. It is distinct from
the paired gauge comparison below.

The additional PT baseline uses ten runs per benchmark instance at a fixed
$0.75$\,s budget. Its reported aggregate hit rate is $0.982$, compared
with $0.981$ for the SA repetitions. A restricted TN repeatability
check ran three anneals on each of the three ungauged Campaign A
Wishart instances at $n=40$. The returned energies were bit-identical
within each instance. This concerns only those three matrices and does
not establish determinism of TN optimisation in general.

\subsection{Effect of planted-optimum placement (Campaign B)}
\label{sec:controls}

The strong Gurobi comparison in Campaign A leaves open whether the fixed
placement of the planted Wishart optimum aids search and hence presents a bias.
The paired control experiment hence tries to isolate that effect. 
Building on that, the enlarged set broadens the instance sample.

\paragraph{Paired optimum-placement control.}
For the Wishart instances, we test whether solver performance depends on the placement of 
the planted optimum, comparing each ungauged instance with a gauge-transformed copy.
Hence, we re-solve the nine Wishart instances
at $n=32,40,48$ both as generated and after random spin reversals.
The gauge transformation described in Sec.~\ref{sec:families} is checked
against exhaustive enumeration at small sizes. This paired comparison
tests sensitivity to solution placement~\cite{MandraKatzgraber2018}.

In the paired control at $n=32$ and $40$, branch-and-cut
proves optimality in both frames. The medians of the six per-instance gauged/plain ratios are
$2.2$ for full solver-wrapper runtime and $1.2$ for node count. At $n=48$, branch-and-cut finds the optimum on all three ungauged instances, but on only one of their gauged counterparts. 
This reduction suggests that the all-zero placement (ungauged) helped branch-and-cut find an optimum on these instances.

\paragraph{Enlarged gauged comparison and TN coverage.}
To extend the paired comparison, the enlarged gauge-randomised set
adds seven additional independent instances at each of
$n=40$ and $48$, giving ten instances per size. 
Branch-and-cut uses
one run per instance with a $600$\,s limit, and SA uses ten runs per
instance (200 runs in total). PT uses five runs per instance at
$0.75$\,s and three at $60$\,s. A separate three-run comparison uses
2000 sweeps (median $1.7$\,s and $2.0$\,s per run) and reaches the
optimum in $29$ of $30$ runs at $n=40$ and $22$ of $30$ at $n=48$.
Gauges are fixed across repetitions. The TN recheck
covers the three Campaign A instances per size, a subset of the enlarged
set, with one anneal in each frame. The restricted TN repeatability check
is described in Sec.~\ref{sec:controlresults}. LR-QAOA is not part of
this paired/enlarged comparison.
The scaling study uses its own instances and budgets
(Sec.~\ref{sec:frontier}). 

On the enlarged gauged set, branch-and-cut reaches all ten optima
at $n=40$ and proves seven (Wilson $95\%$ interval:
$0.40$--$0.89$), while SA reaches the optimum on $0.63$ of runs
($0.53$--$0.72$). At $n=48$, branch-and-cut reaches four of ten
($0.17$--$0.69$) and proves none, while SA reaches $0.25$
($0.18$--$0.34$). Intervals for run-level rates such as SA's pool the
runs on these fixed instances. They describe how the solver performs on
these instances and do not account for variation between instances, so
they understate the uncertainty for new instances from the same ensemble.
The Gurobi and SA results above cover all ten gauged instances at each size. 
The TN comparison covers only the three instances per size inherited from Campaign A. 
Campaign E uses a separate instance set, described in Sec.~\ref{sec:frontier}.

PT reaches every optimum in every tested $60$\,s run on the enlarged
gauged set (Table~\ref{tab:taxonomy}). TN reaches one optimum in its
six-instance gauged subset, as in the six ungauged Campaign A
runs. The comparison separates search from certification: PT can find
optima that Gurobi misses within its budget, while Gurobi alone supplies
certificates on seven $n=40$ instances.

\begin{table}[tbp]
\caption{Enlarged gauged Wishart comparison (Campaign B). Gurobi uses
one run per instance with a $600$\,s cap. SA uses ten runs per instance.
PT uses three runs per instance at $60$\,s for the results shown.
TN covers only the three Campaign A instances per size, not the full
ten-instance set. All references are planted optima. LR-QAOA is not
part of this comparison.}
\label{tab:taxonomy}
\begin{tabularx}{\textwidth}{@{}lYYYY@{}}
\toprule
$n$ & Gurobi: hits; certificates & SA: fraction of runs & PT at $60$\,s: fraction of runs & TN: subset hits \\
\midrule
40 & $10/10$; $7/10$ & $0.63$ & $1.00$ & $1/3$ \\
48 & $4/10$; $0/10$ & $0.25$ & $1.00$ & $0/3$ \\
\bottomrule
\end{tabularx}
\end{table}
\section{Tensor-network diagnostics, failure mechanisms, and solver selection}
\label{sec:regionT}

We first analyse measurements from Campaign A, then use the targeted
experiments in Campaign C to investigate failure mechanisms. These
analyses distinguish representation demand, runtime, and search success.

\subsection{Structural prediction of representation demand and runtime}

\begin{figure}[tbp]
  \includegraphics[width=\linewidth]{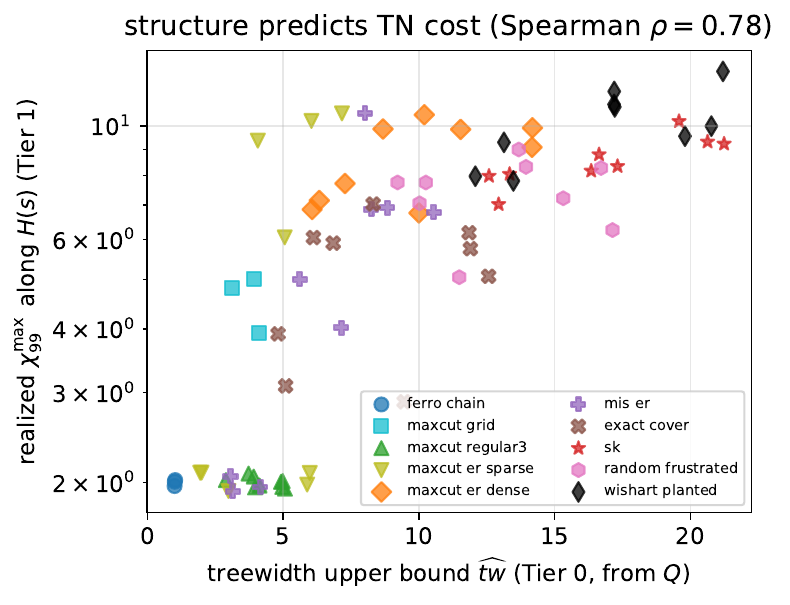}
  \caption{Structure predicts tensor-network cost: Tier-0 treewidth
  bound vs.\ realised bond-dimension demand $\chimax$ along $H(s)$
  (calibration set, 78 distinct matrices with verified optima; Spearman
  $\rho = 0.78$, and $\rho = 0.73$ on the out-of-sample set at
  $n = 32$--$48$). This
  regression is the TN leg of the cost model, Eq.~\eqref{eq:ttn}.}
  \label{fig:h3}
\end{figure}

The bond-dimension demand $\chimax$ along the TN trajectory correlates
with the $O(n^3)$ treewidth bound at $\rho = 0.78$ on the calibration
set and $0.73$ out of sample, and with the cheaper RCM
bandwidth at $\rho = 0.72$ and $0.67$ (Fig.~\ref{fig:h3}). These values
are computed over the 78 distinct matrices of each split
(Sec.~\ref{sec:split}); counting the repeated grid and ferro-chain
matrices three times would raise them to $0.81$ and $0.76$. This is a
reliable ranking, not a precise cost
estimate: the regression explains $55\%$ of the variance in
$\log_2 \chimax$ on the distinct calibration matrices ($R^2 = 0.55$). With measured $\chimax$, the runtime
model predicts wall-clock to a median factor of $1.8$ out of sample.
Using treewidth alone, the genuinely a-priori median factor is $7.6$
(Sec.~\ref{sec:advisor}). It therefore supports coarse cost comparisons.

That predictability does not imply that TN is fast. Among the $44$ instances
whose trajectory needs only $\chi_{99} \le 2$, its anneal is a median
$17\times$ slower than SA. At $n = 48$ the median gap is $1600\times$.

\begin{figure}[tbp]
  \includegraphics[width=\linewidth]{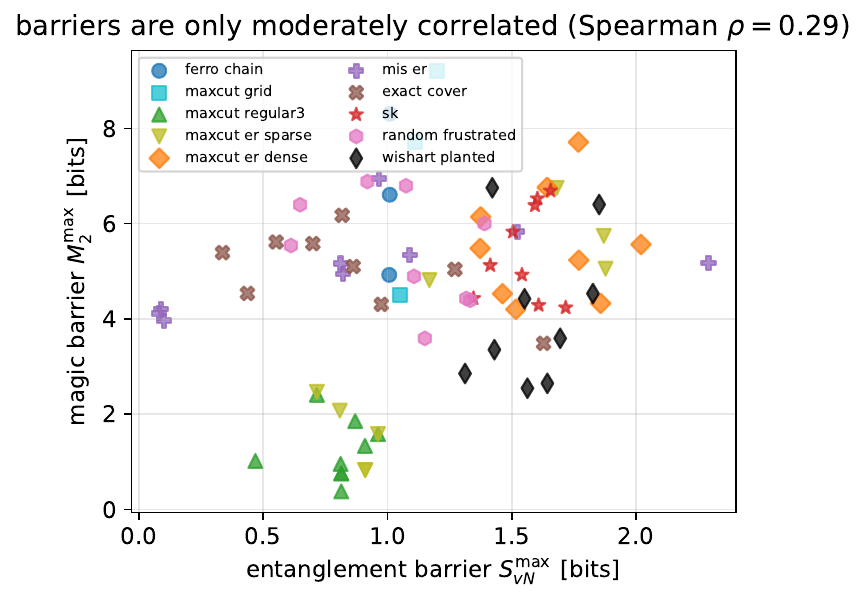}
  \caption{  The two trajectory resources are only moderately correlated: magic
  barrier $\Mmax$ vs.\ entanglement barrier $\Smax$ over the 78 distinct calibration matrices (Spearman $\rho = 0.29$ here, $0.45$ out of sample). This
  supports treating them as complementary, rather than interchangeable,
  diagnostics of the computed states.}
  \label{fig:h2}
\end{figure}

Entanglement and magic are only moderately associated in the computed diagnostic
trajectories ($\rho=0.29$ in calibration and $0.45$ out of sample over
distinct matrices, see also
Fig.~\ref{fig:h2}). They provide complementary descriptions of these
states, rather than interchangeable measures of representation demand.

\subsection{Capacity sensitivity, trapping, and reset mitigation (Campaign C)}
\label{sec:tnmitigation}

Modest measured representation demand does not ensure successful
variational optimisation. Campaign C compares reset-enabled solver runs
at $\chi=4,8,16$ and examines selected trajectories against independently
computed spectra. These runs differ from the reset-free diagnostic
sweeps in Sec.~\ref{sec:tier1}. Detailed spectral settings and the
capacity-failure breakdown are given in App.~\ref{app:tnsupport}.

\begin{figure}[tbp]
  \includegraphics[width=\linewidth]{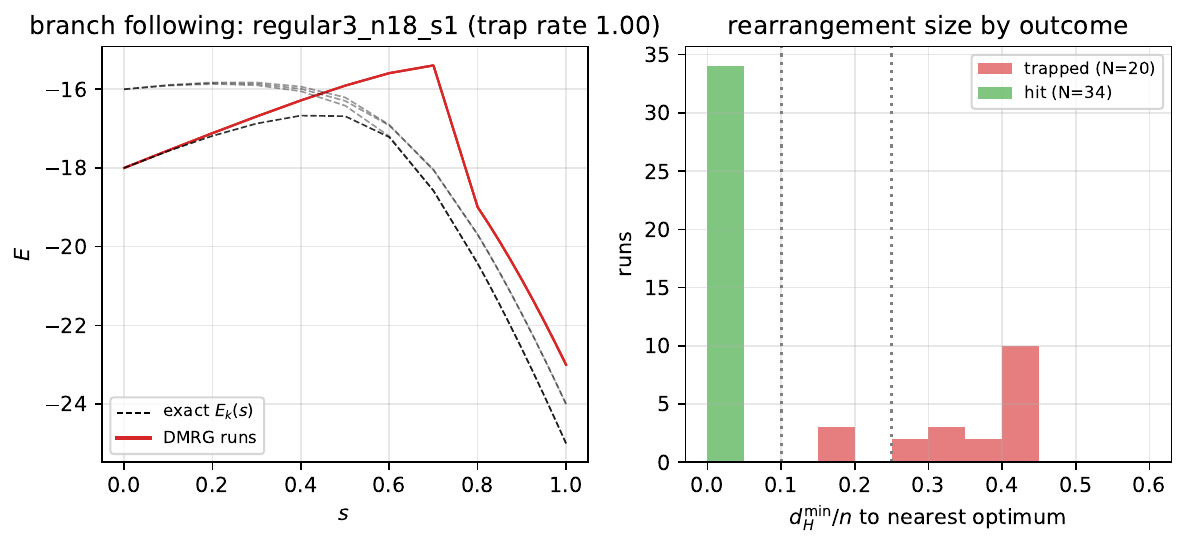}
  \caption{The TN failure mode is dynamical, not capacity-theoretic.
  Left: DMRG energy traces of six anneals on the worst 3-regular
  instance against the exact levels $E_k(s)$. Trapped runs follow an
  excited branch to $s = 1$. Right: distance to the nearest certified
  optimum by outcome. Trapped runs end extensively far (median
  $d_H^{\min}/n = 0.40$), successful runs at $d_H = 0$.}
  \label{fig:trapping}
\end{figure}

The targeted trajectories identify excited-state trapping as a failure
mechanism. A warm-started sweep initialises each $s$-point
from the preceding converged state and can follow an excited spectral
branch through an avoided crossing instead of the branch that becomes
the ground state. On 3-regular Max-Cut, trapped runs end in excited
computational-basis states at a median normalised distance
$d_H^{\min}/n = 0.40$ from the nearest optimum, whereas successful runs
end at an optimum (Fig.~\ref{fig:trapping}).

We tested a crossing-detection reset: at checkpoints in the second half
of the sweep, we launch a cold-started DMRG from a fresh MPS and replace
the warm-started state if the new state has lower energy. This cures all previously
resistant cases with one anneal: over the eight evaluated instances
($3$-regular Max-Cut and ferromagnetic controls at $n = 18$ and $22$) the
reset reaches the optimum on $8$ of $8$, against $5$ of $8$ for the
baseline and $6$ of $8$ for four-fold multi-seeding, and the
cost-optimised variant requires exactly one reset per cure. With trapping controlled, capacity failures are rare: among
the $81$ mitigation-clean out-of-sample instances, one fails at
$\chi = 4$ and none at $\chi = 16$. The remaining failures are not nested in bond cap, consistent with
residual trapping rather than a simple capacity limit. The diagnostic
$\chimax$ does not meet the stated discrimination target. Detailed counts
and AUC values are given in App.~\ref{app:tnsupport}.

\subsection{Limits of predictive solver selection}
\label{sec:advisor}

To assess whether runtime prediction supports heuristic selection, we fit
cost models on the 90 calibration instances and evaluate them on the 90
larger instances from Campaign A (App.~\ref{app:costmodels}). The accuracy
summarised above concerns runtime. Selecting a solver additionally
requires information about its probability of reaching the target.

These cost predictions do not provide a useful selection rule. The
cost-based ranking matches the single-run ex-post heuristic label on $85$
of $90$ out-of-sample instances, but an always-predict-SA baseline scores
identically and the balanced accuracy is $0.50$. 
The ranking never selects TN. 
Moreover, the five apparent misses are Wishart benchmark instances from Campaign A with single-run labels, 
while repeated SA and PT runs show that those isolated observations are not stable solver properties.
The model predicts
runtime, not the probability of reaching the target. Success-probability
weighting does not resolve this limitation (App.~\ref{app:costmodels}).

Runtime prediction provides limited information about cost, but does not
supply the success-probability information needed for a useful heuristic
selection rule on these instances.

\section{Quantum comparison and limits of extrapolation}
\label{sec:regionQ}

The main LR-QAOA comparison belongs to Campaign A. Campaign D tests
adaptive-depth coverage, larger-instance simulation convergence, and
small-size success-probability scaling on separate instance sets.

\subsection{Main LR-QAOA comparison and supplementary checks}
\label{sec:circuitchecks}

LR-QAOA produces no solution-quality win in the 105 simulation runs: it
never beats the better of SA and the TN anneal. These comprise 30 exact
statevector runs at $n \le 16$ and 75 MPS-proxy runs at $18 \le n \le 32$. It underperforms on exact cover
($3/9$ hits), MIS ($9/12$), Wishart ($5/9$) and one dense-ER instance
($11/12$), while matching the classical result on the other families.
The adaptive-depth check covers all 27 benchmark instances at actual
$n=32$, yielding $3/12$ hits on newly covered instances and $13/15$ on
previously covered instances (App.~\ref{app:circuitsupport}).
Simulation wall-clock time measures classical computational cost.
Quantum-circuit depth is a separate hardware-relevant resource estimate,
not a complete measure of quantum runtime~\cite{Dehn2026}. This is a negative result for the
fixed-schedule protocol tested, not for all QAOA variants, since angle
optimisation, warm starts and recursive methods are outside scope.

A separate MPS simulation-convergence test uses six gauge-randomised Wishart
instances at $n=40,48$. All tested bond caps saturate, sampled energies
do not converge as the cap increases, and no run reaches the optimum.
A deeper probe is available for only one instance. These runs therefore
do not provide a converged LR-QAOA simulation on the dense larger
instances (details are in App.~\ref{app:circuitsupport}).

\subsection{Small-size success-probability scaling (Campaign D)}
\label{sec:probscaling}

An LR-QAOA layer count proportional to size would give polynomial
gate-level circuit depth on
all-to-all connectivity, but an efficient solver also requires adequate
success probability. We test that premise at small sizes. The
conditional resource calculation and exploratory candidate indicator
are given in App.~\ref{app:quantumexploration}.

\begin{figure}[tbp]
  \includegraphics[width=\linewidth]{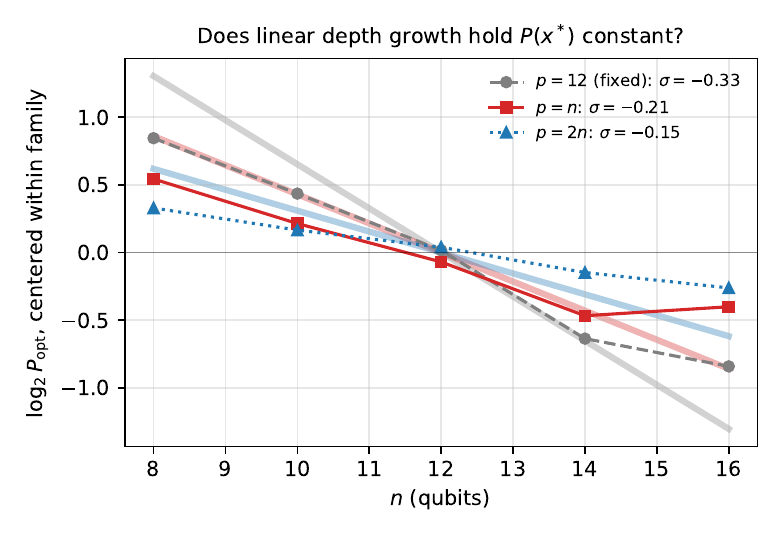}
  \caption{Testing the premise the depth extrapolation imports: exact
  statevector LR-QAOA success probability $P_{\mathrm{opt}}$ against
  instance size, at fixed depth $p = 12$ and at linearly growing depths
  $p = n$ and $p = 2n$. Family-centred values; faded lines are the fitted slopes. If linear depth growth held
  $P_{\mathrm{opt}}$ constant, the $p = n$ curve would be horizontal;
  all three decline.}
  \label{fig:depthscaling}
\end{figure}

We compute the exact probability mass on the full optimal manifold
for $n=8$--$16$, across ten families and three seeds, at $p=12,n,2n$.
The 450 exact statevector measurements evaluate probability mass rather
than the polished sampled energies used in Campaign A.
The fitted slopes of $\log_2 P_{\mathrm{opt}}$ are $-0.32,-0.21,-0.15$
bits per qubit, respectively: increasing depth slows the decay, but at
$p=n$ the probability still halves about every five qubits. Against the
pre-registered criterion, which counted $|\sigma| \le 0.1$ as support and
$\sigma \le -0.25$ as refutation, this result is inconclusive, not a
confirmation of constant success probability. The family breakdown is consequential:
six families are near-flat, but the slopes at $p=n$ remain negative for exact
cover ($-0.75$), MIS ($-0.54$) and Wishart ($-0.29$), including the family
proposed as the first quantum target. All three lie below the refutation
threshold of $-0.25$. The criterion was registered for the pooled slope,
so this family-level reading is descriptive, but it points against the
premise on Wishart rather than leaving it open. The quantum-candidate region and
its polynomial-depth interpretation must therefore remain conditional:
our small-size data do not support the required scaling premise on
Wishart. The measurement window is narrow, however, so it cannot rule out
asymptotic softening of the decay rate. Figure~\ref{fig:depthscaling}
shows the decline across the tested depth schedules.

\section{Wishart hardness--size--budget scaling}
\label{sec:frontier}

Campaign E extends the gauged Wishart comparison to larger sizes and a
second hardness setting. The planted optimum supplies a reference
independent of solver success, allowing search and certification to be
tested directly rather than inferred from structural diagnostics.

\subsection{Scaling protocol (Campaign E)}

Because the raw optimum is a special bitstring
found quickly by Gurobi's root-node primal heuristics, we gauge-randomise each instance to
move it to a generic bitstring while preserving the energy landscape up to
spin relabelling. 
We vary the hardness parameter $\alpha = M/n$, where $M$ is the number of planted
Gaussian sample vectors \cite{Hamze2020wishart}, at $\alpha = 0.5$ and
$0.7$ (lower is harder). The native binary-quadratic formulation is solved
with a $600$\,s default limit and one-hour limits on selected instances below.
Campaign E includes 67 exact-solver runs on 61 distinct instances; 
six runs repeat an instance with a larger budget. Exact-solver summaries
use one run result per instance, favouring successful or larger-budget runs.

SA uses ten seeds per instance at its fixed update budget. PT uses three
seeds per instance at each of $1$, $6$, and $60$\,s. Six instances solved
in all three six-second PT runs were not repeated at 60 seconds.
For these six instances, the reported 60-second results reuse the six-second
success rates (no additional 60-second runs were performed).
For six instances ($\alpha=0.5$, $n=48$ and $64$), the stored SA and PT
logs contain a second batch run with the same seeds. These replays are
not additional replication, so only the first batch per instance is
counted.

\subsection{Search success across hardness, size, and budget}

\begin{figure}[tbp]
  \centering
  \includegraphics[width=0.90\linewidth]{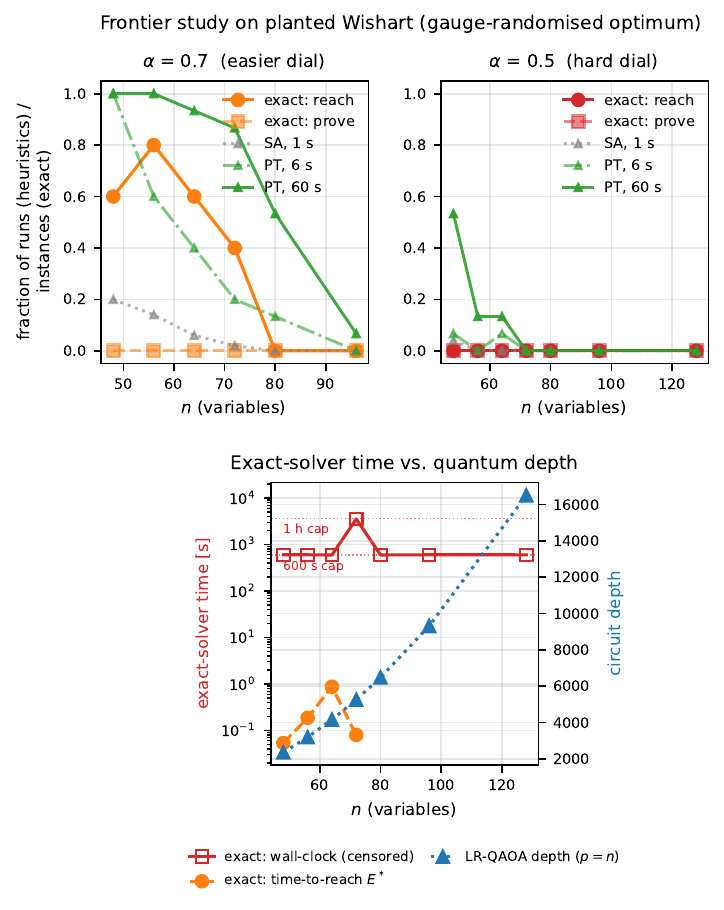}
  \caption{Frontier scaling on the gauge-randomised planted Wishart
  ensemble. Five instances per cell where available; licence-token
  outages reduced some counts. Top row: rates of reaching the optimum by
  hardness parameter $\alpha$. Exact-solver rates are per instance, heuristic rates per
  run (SA ten runs per instance, PT three). Bottom: exact-solver
  wall-clock (open red symbols) is censored at the budget. Time-to-reach
  (orange) includes only successful runs, and the blue curve gives LR-QAOA
  depth at $p=n$.
  The exact-solver limit is $600$\,s except for one-hour runs on seeds
  $0,1,2$ at $\alpha=0.5$, $n=48,64$, and all five instances per
  aspect ratio at $n=72$. Replayed heuristic batches with identical
  seeds ($\alpha=0.5$, $n=48,64$) are counted once. For the hard instances ($\alpha=0.5$)
  no tested classical method finds the optimum from $n=72$ onward. For the
  easier instances ($\alpha=0.7$), heuristic reach extends with budget. The polynomial-depth
  comparison remains conditional on the unconfirmed $p\sim n$ premise.}
  \label{fig:frontier}
\end{figure}

Table~\ref{tab:ladder} compares heuristic reach rates across budgets and
hardness settings. Fig.~\ref{fig:frontier} shows the corresponding exact-
solver results and timing.

\begin{table}[tbp]
\caption{Frontier scaling study: the rate at which a solver reaches the
gauge-randomised planted optimum, using ten SA seeds and three PT seeds
per instance; replayed batches with identical seeds are counted once.
At $\alpha=0.7$ (easier), increasing the PT budget extends the range of instance sizes at which optima are found. 
At $\alpha=0.5$ (harder), no tested budget yields an optimum at the sampled sizes from $n=72$ onward.
The SA column uses its fixed update budget (approximately one second),
not a wall-clock stopping rule. PT results at 60 seconds include instances
already solved in every run at six seconds without repeating them.
$^{*}$marks the cell entirely inherited from the six-second results.}
\label{tab:ladder}
\begin{tabularx}{\textwidth}{@{}l*{5}{>{\centering\arraybackslash}X}@{}}
$\alpha$ & $n$ & SA $\sim1$\,s & PT $1$\,s & PT $6$\,s & PT $60$\,s \\
\midrule
$0.7$ & $48$ & $0.20$ & $0.67$ & $1.00$ & $1.00^{*}$ \\
      & $56$ & $0.14$ & $0.27$ & $0.60$ & $\mathbf{1.00}$ \\
      & $64$ & $0.06$ & $0.07$ & $0.40$ & $\mathbf{0.93}$ \\
      & $72$ & $0.02$ & $0.07$ & $0.20$ & $\mathbf{0.87}$ \\
      & $80$ & $0$    & $0$    & $0.13$ & $\mathbf{0.53}$ \\
      & $96$ & $0$    & $0$    & $0$    & $0.07$ \\
\midrule
$0.5$ & $48$ & $0.04$ & $0$    & $0.07$ & $\mathbf{0.53}$ \\
      & $56$ & $0$    & $0$    & $0$    & $0.13$ \\
      & $64$ & $0$    & $0.07$ & $0.07$ & $0.13$ \\
      & $72$--$128$ & $0$ & $0$ & $0$ & $\mathbf{0}$ \\
\end{tabularx}
\end{table}

The table shows that the search wall depends on both hardness and budget:
on $\alpha = 0.7$, longer PT runs reach larger instances, while on
$\alpha = 0.5$ no tested classical method reaches the optimum from
$n = 72$ upward. At those sizes this is based on $95$ heuristic runs per
size at $n = 72, 80, 96$ and $19$ runs on the single surviving
$n = 128$ instance. 

\subsection{Certification and longer-budget exact-solver runs}

Across all $61$ gauge-randomised instances with
$n \ge 48$, branch-and-cut certifies none (median mixed-integer-programming (MIP) gap
${\approx}\,0.2$). For the harder instances ($\alpha=0.5$) it also finds none of the $31$
instances. One-hour runs at $n = 48, 64, 72$ confirm the misses, despite
exploring up to ${\sim}5\times10^{7}$ nodes. For the easier instances ($\alpha=0.7$), exact
solver hits occur early, all within $2.3$\,s (median $0.1$\,s), and disappear by
$n = 80$. All recorded SA hits with $\alpha=0.5$ occur at $n=48$.

The one-hour limit was used at $\alpha=0.5$ for three seeds
at $n=48,64$, and for all five instances at $n=72$ at both aspect
ratios. The remaining instances retain the $600$\,s cap. Recurring Gurobi licence-token outages truncated
some long runs and prevented broader escalation. These timeouts do not
prove that the optimum is unreachable in principle, so shorter-budget
results are weaker evidence. The failure at $\alpha=0.5$ identifies a candidate
regime for a hardware experiment, not a quantum win: polynomial depth
still depends on the unconfirmed $p \sim n$ success-probability premise in
Sec.~\ref{sec:probscaling}, and a classically simulable circuit would not
outperform its own simulation.

\section{Discussion}
\label{sec:discussion}

The campaigns distinguish search, certification, and diagnostic
explanation. Campaign A establishes a strong branch-and-cut baseline:
it matches every reference energy and certifies 176 of 180 benchmark
instances. Campaign B qualifies that comparison. On the enlarged gauged
Wishart set, PT reaches optima that branch-and-cut misses within its
budget, although it cannot provide certificates. The preferred method
therefore depends on the required outcome and the instance setting,
not only on the variable count.

The Campaign A diagnostics rank TN representation demand and provide a
coarse runtime estimate, but do not yield a useful solver-selection
rule. Campaign C identifies excited-state trapping and demonstrates
reset mitigation on the evaluated cases. Among the 81 mitigation-clean
out-of-sample instances, one fails at $\chi=4$ and none at $\chi=16$.
However, failures remain in the full 90-instance out-of-sample set, 
so the results on the 81 mitigation-clean instances do not establish 
that representation-capacity limits or optimisation failures have been eliminated in general.

The 105 LR-QAOA simulation runs in Campaign A show no solution-quality improvement
over the classical comparators. Campaign D extends coverage and tests
simulation and probability-scaling assumptions, without establishing an
efficient quantum solver on Wishart. Campaign E supplies a separate
classical scaling comparison: at $\alpha=0.5$, none of the tested
methods reaches the planted optimum at sampled sizes from $n=72$
onward within the tested budgets. This identifies a setting for further
comparison, rather than evidence of quantum advantage.

\subsection{Threats to validity}
\label{sec:threats}

The heuristic comparison depends on the solver portfolio and budget.
In Campaign B's enlarged gauged set at $n=48$, the reported hit rate
rises from SA's $0.25$ to PT's $0.44$ at a $0.75$\,s allocation and to
$1.00$ at $60$\,s per PT run. In Campaign E, increasing the PT budget
also extends search success on $\alpha=0.7$ (Table~\ref{tab:ladder}).
These comparisons concern the tested implementations and the budgets
specified for each campaign, not heuristic optimisation in general.
A portfolio solver, tabu search, or Burer--Monteiro relaxation could
change the comparison further.

The Wishart generator places its planted optimum at $x = 0$, the trivial
incumbent of the mixed-integer program. On those instances the exact
solver starts at the answer and only the bound is
work, which flatters it. SA is affected too, through its temperature
schedule. Its start temperature $T_0=\max_i\sum_j|Q_{ij}|$ reads the
diagonal of $Q$, which the gauge transformation changes, whereas its
other steps (uniform random start, Metropolis acceptance on energy
differences) are gauge-equivariant. In a control with ten independent
runs on each of the ten instances per size, SA's rate at $n = 48$ falls
from $0.44$ on the planted instances to $0.25$ on the gauged ones (exact
sign test over instances, $p = 0.008$). Running the gauged instances with
the planted frame's $T_0$ restores it to $0.39$. At $n = 40$ the frames
do not differ ($0.63$ in both). Part of SA's gauged deficit at $n = 48$
therefore reflects its schedule rather than the landscape.
Campaign A uses ungauged Wishart instances (Table~\ref{tab:outcomes}).
Campaign B distinguishes paired and enlarged gauged comparisons
(Table~\ref{tab:taxonomy}), and Campaign E uses gauged instances
throughout. The separate paired gauging control used three instances per size. At
$n = 48$, the exact solver's reach in that small sample falls from $3$ of
$3$ on planted instances to $1$ of $3$ after gauge transformation.

SA's Campaign A one-hit-in-six result is a single-run summary on the
ungauged Wishart instances. Ten additional runs per instance give a
pooled hit rate of $0.53$. TN repeatability was checked only on a
restricted ungauged subset and does not establish general repeatability
(Sec.~\ref{sec:controlresults}). Likewise, SA's 900/900 calibration hits do not justify
assuming that single runs are representative on harder instances.

The enlarged gauged comparison in Campaign B uses ten instances per
size, while its TN recheck covers three. Campaign E has its own sample
counts, including only one instance at $n=128$. The denominators and
replication therefore differ across claims. For example, the LR-QAOA
comparison's exact-cover hit count is $3/9$. App.~\ref{app:repro}
summarises replication. These limited samples and the reporting issues
noted with the protocols constrain generalisation beyond the tested
instances.

Wall-clock comparisons reflect implementation, threading, and system
load as well as algorithmic work. Gurobi uses compiled solver code.
SA and PT use Python update loops, while TN uses compiled numerical
routines through Python. Shared language alone does not remove these
differences. Timing boundaries also differ: the Campaign A exact baseline
times optimisation, heuristic run times include the full solver
call, and PT excludes replica initialisation. These differences limit
the interpretation of cross-solver runtime ratios.

Campaign A's MPS simulation of LR-QAOA is limited by fixed depth and the
gate-count guard. Campaign D improves coverage at $n=32$ by changing
depth. Its newly covered dense cases use $p=6$--$11$. The separate
$n=40,48$ probes do not establish convergence of the truncated MPS
simulation of LR-QAOA (App.~\ref{app:circuitsupport}). These limitations concern
the tested simulations and do not establish LR-QAOA performance on
hardware. The success-probability scaling premise is also unconfirmed
in the measured window (Sec.~\ref{sec:probscaling}).

Our LR-QAOA simulations do not model device noise, and the circuit-depth
estimates do not by themselves establish performance on noisy hardware.
Stilck Fran\c{c}a and Garc\'ia-Patr\'on~\cite{StilckFranca2021}
derive limits on noisy quantum optimisation that depend on the noise
level and the relation between problem and device connectivity.
Their results motivate treating noise and connectivity explicitly in
a hardware comparison. We do not evaluate those bounds for the
instances studied here.

The Campaign A calibration and evaluation sets differ structurally because the split by
size was intentional for the extrapolation question, but it means the two
sets are not exchangeable. SA reaches the optimum in every calibration run,
in $96\%$ of out-of-sample runs, and on Wishart out-of-sample in $66\%$.
A quality model fitted where nothing fails cannot learn what failure looks like,
which is exactly the obstruction App.~\ref{app:costmodels} runs into.

Campaign A covers ten synthetic families up to $n=48$. Its solver
comparison need not transfer to industrial instances or other size
ranges. Campaign E extends the evidence to larger Wishart instances,
but its failures remain statements about tested budgets, not proofs of
intractability. The default exact-solver cap is ten minutes, with
one-hour limits on selected instances as specified in
Sec.~\ref{sec:frontier}. Fixed-budget comparisons alone do not establish
asymptotic advantage~\cite{Ronnow2014}.

The conditional quantum-resource projection in
App.~\ref{app:quantumexploration} also depends on family-specific
success-probability scaling. The cited LR-QAOA study reports favourable
behaviour for fully-connected weighted Max-Cut, but not its Max-2-SAT
family~\cite{MontanezBarrera2025}. Campaign D does not establish the
required premise for Wishart in the measured small-size window.
Campaign E supplies classical evidence only. A direct quantum test on
the larger gauged Wishart instances remains necessary.

Finally, the evaluated success-probability weighting does not improve
the Campaign A cost ranking (Sec.~\ref{sec:advisor}). Campaign B shows
why isolated SA runs cannot support stable solver recommendations, and
why PT belongs in the comparison. 

\section{Conclusions}
\label{sec:conclusions}

The comparison of QUBO solvers depends on whether the objective is to
find a solution or to certify its optimality. In Campaign A, with 180
benchmark instances, general-purpose branch-and-cut matches every
reference energy and proves optimality on 176 instances within the
$600$\,s cap, including 86 larger instances outside the enumeration
range used here. This establishes a strong exact-solver baseline for
the tested families and sizes, rather than a universal recommendation
based on variable count.

The gauge-randomised Wishart controls in Campaign B show why the distinction between
search and certification is essential. At $n = 40$, branch-and-cut
certifies seven of ten optima. At $n = 48$, it finds four of ten and
certifies none within its budget. PT, given $60$\,s per run, reaches
every optimum in every tested run on this enlarged control set, but
provides no certificate. Thus the method that supplies proofs need
not be the most effective method for finding solutions within a
given budget.

The Campaign A trajectory diagnostics provide a cost estimate, and
Campaign C provides an explanation of a solver failure. A treewidth bound ranks the
tensor-network trajectory's bond-dimension demand, although a tighter
runtime estimate requires the measured bond dimension. Controlled
experiments show that warm-started DMRG can follow an excited branch
and that resets resolve the failures in the targeted mitigation study.
Failures remain in the broader out-of-sample evaluation, however, so
the reset procedure does not eliminate trapping in general.
Entanglement and magic provide distinct,
moderately correlated descriptions of the measured trajectories.
These results support diagnostic use of the features, but the
evaluated cost models do not improve solver selection over the
always-SA baseline.

The tested LR-QAOA runs show no solution-quality advantage over the
classical comparators. Campaign E extends the
classical comparison to $n = 128$: on the harder setting
($\alpha = 0.5$), none of the tested classical methods reaches the
planted optimum at the sampled sizes from $n = 72$ onward.
This is a result for the tested implementations and budgets, with
one-hour exact-solver runs on selected sizes, not a bound on
classical optimisation in general. It identifies a concrete regime
for further comparison without establishing a quantum advantage.
In particular, the small-size measurements in Campaign D do not establish the
success-probability scaling needed to make LR-QAOA efficient on
Wishart instances. A direct quantum evaluation in this regime,
together with stronger classical baselines and estimates of
success probability at comparable computational budgets, is needed
to determine whether an advantage exists.

\section*{Abbreviations}
\begin{tabular}{@{}ll@{}}
AUC & area under the ROC curve \\
BF & brute-force enumeration \\
DMRG & density-matrix renormalization group \\
ER & Erd\H{o}s--R\'enyi \\
LR-QAOA & linear-ramp quantum approximate optimization algorithm \\
MIP & mixed-integer programming \\
MIS & maximum independent set \\
MPS & matrix-product state \\
PT & parallel tempering \\
QAOA & quantum approximate optimization algorithm \\
QUBO & quadratic unconstrained binary optimization \\
RCM & reverse Cuthill--McKee \\
ROC & receiver operating characteristic \\
SA & simulated annealing \\
SK & Sherrington--Kirkpatrick \\
SVD & singular-value decomposition \\
TN & tensor network \\
\end{tabular}

\section*{Declarations}

\bmhead{Availability of data and materials}
The datasets and code supporting the conclusions of this article are
deposited in the Zenodo repository,
\url{https://doi.org/10.5281/zenodo.23661253}~\cite{ZenodoRelease}. Access is
restricted until publication of this article and has been provided to the
editor and reviewers; the archive will be made publicly available upon
publication. It contains the per-run result records of all campaigns, the
code that produced them and a script that recomputes the numbers reported
in this paper; the computing environment is documented in
App.~\ref{app:repro}. Project name: qubo-advisor. Project home page and
archived version: \url{https://doi.org/10.5281/zenodo.23661253}. Operating
system: Linux. Programming language: Python~3.12. Other requirements:
NumPy, SciPy, networkx, matplotlib, quimb; Gurobi for the exact-solver
runs. Licence: MIT (code), CC~BY~4.0 (data). Restrictions to use by
non-academics: none; Gurobi requires its own licence.

\bmhead{Ethics approval and consent to participate}
Not applicable. This study involves neither human participants nor animals.

\bmhead{Consent for publication}
Not applicable.

\bmhead{Competing interests}
The authors declare that they have no
competing interests.

\bmhead{Funding}
Open Access funding enabled and organized by Projekt
DEAL. This work was supported by the Ministry of Economic Affairs, Labour
and Tourism Baden-W\"urttemberg in the frame of the Competence Center
Quantum Computing Baden-W\"urttemberg (project no.\ KQCBW25).

\bmhead{Authors' contributions}
All authors contributed to the design of
the study, the implementation, the analysis and the writing, and read
and approved the final manuscript.

\bmhead{Acknowledgements}
Not applicable.

\begin{appendices}
\section{Reproducibility}
\label{app:repro}

All wall-clock figures quoted in this paper were measured on one machine:
an Intel Core Ultra 9 285H (16 logical
cores) with $15$\,GB of RAM available to a WSL2 Linux guest (kernel
6.6.87), running Python 3.12.12 with NumPy 2.3.5, SciPy 1.16.3, quimb
1.11.2 and networkx 3.6.1. The exact solver is Gurobi 13.0.2.
The code base records this environment in machine-readable form.
Implementation differences across the solvers are discussed in
Sec.~\ref{sec:solverdesc}.

Main-benchmark budgets and replication are given in
Table~\ref{tab:budgets}. Campaign B protocols are specified in
Secs.~\ref{sec:controlresults} and~\ref{sec:controls}. Campaign D protocols are given in
Sec.~\ref{sec:probscaling} and App.~\ref{app:circuitsupport}.

Campaign E contains
61 distinct instances and 610 distinct SA instance/seed combinations.
Its stored SA and PT repeat files also contain six replayed batches
with identical seeds; all reported rates count only the first batch
per instance (Sec.~\ref{sec:frontier}). The treatment of PT results
carried forward from shorter budgets is specified in the same section.

\section{Solver configuration details}
\label{app:solvers}

This appendix gives the numerical settings and implementation checks
for the solvers introduced in Sec.~\ref{sec:solverdesc}.

\paragraph{Exact solver.}
We minimise $x^TQx$ over binary variables using
Gurobi~\cite{Kochenberger2014,Gurobi}, with either the standard product
linearisation or the native binary-quadratic objective. The saved main
baseline uses 165 linearised and 15 native quadratic models. The latter
are the SK and Wishart instances at $n=40,48$ and the random-frustrated
instances at $n=48$, whose earlier linearised attempts encountered the
solver's licence-size limit. The linearised objective is checked against
exhaustive enumeration at small sizes. For each instance, we use the
result from the final successful solver invocation.

At termination, the baseline records the best feasible solution,
remaining MIP gap, solver status, and elapsed optimisation time.
Reaching the reference energy and certifying optimality are separate
outcomes. The certificate flag means that Gurobi returned
\texttt{OPTIMAL} under its numerical tolerances, with the default
relative MIP-gap tolerance of $10^{-4}$. Four of the 176 flagged
records have nonzero relative MIP gaps, ranging from
$8.66\times10^{-7}$ to $9.06\times10^{-5}$; all lie below this
tolerance, and we count them as certified.

As an upper limit, these runs are capped at $600$\,s.

\paragraph{Simulated annealing.}
SA~\cite{Kirkpatrick1983} uses single-flip Metropolis updates with
$R=4$ restarts and $n_{\mathrm{sw}}=2000$ sweeps per restart.
A sweep proposes one flip at each of the $n$ sites in random order,
giving $8000n$ proposed updates per run. The geometric temperature
schedule starts at $T_0=\max_i\sum_j|Q_{ij}|+10^{-12}$ and ends at
$T_1=10^{-3}$, with ratio $(T_1/T_0)^{1/(n_{\mathrm{sw}}-1)}$.
Because $T_0$ includes the diagonal of $Q$, it is not invariant under
the gauge transformation; Sec.~\ref{sec:threats} quantifies the effect.
The best state from each restart is refined by first-improvement
descent, and the best result is returned. Reference-energy generation
also uses $R=8$ and $n_{\mathrm{sw}}=4000$, giving four times as
many proposed updates.

\paragraph{Parallel tempering.}
PT uses 16 replicas on a geometric temperature ladder. Its hot endpoint
is $T_{\max}=\max_i\sum_j|Q_{ij}|+10^{-12}$. Its cold endpoint is
$T_{\min}=\max(\min_{i\ne j,Q_{ij}\ne0}|Q_{ij}|/10,10^{-6})$.
For a matrix without off-diagonal couplings, $T_{\max}/100$ replaces
the inner minimum before applying the floor. Each replica performs
random-order single-flip Metropolis updates. After every sweep, all
neighbouring replica pairs attempt exchanges sequentially. No final
local-search polish is applied. Repetition seeds are $1000(k+1)$.

The clock starts after replica initialisation and is checked after
each complete sweep and exchange pass, permitting a small overshoot.
A time budget supersedes the ordinary sweep limit. Main-benchmark
runs use a fixed $0.75$\,s allocation rather than matching each
instance's measured SA time. Control budgets and repetitions are
specified in Sec.~\ref{sec:controls}.

On the gauged Wishart instances at $n=40$ and $48$, with $0.75$\,s per
run, the median swap acceptance was $0.08$ with eight replicas and
$0.27$ with 16; with eight, too few exchanges are accepted for the
ladder to work, so all reported PT runs use 16. Acceptance in other
runs varies with instance and budget.
The stored PT records do not include the replica count. For the
Campaign A and B results it is fixed by the labelled result files; for
Campaign E it rests on the documented run command (\texttt{--replicas 16}).

\paragraph{Tensor-network anneal.}
The TN solver uses two-site DMRG on a matrix-product-operator
representation of $H(s)$~\cite{White1992,Schollwoeck2011}, checked
against an independent dense construction at $n=6$. It uses an
11-point interpolation grid, with more points near $s=1$, and
warm-starts each point from the preceding MPS. A cold-start reset
may replace that state when it finds a lower energy. The mitigation
study is described in Sec.~\ref{sec:tnmitigation}.
The primary TN solver uses $\chi=32$ on all 180 benchmark instances. Its energies
and wall-clock times supply the main comparison and runtime-model fits.
Separate runs at $\chi=4,8,16$ provide the capacity study. All four
solver configurations enable reset checks, using the common grid
\[
s\in\{0,0.16,0.32,0.48,0.64,0.80,0.84,0.88,0.92,0.96,1\}.
\]
DMRG uses truncation cutoff $10^{-10}$, convergence tolerance $10^{-8}$,
and at most 30 sweeps per point. At every grid point with $s\ge0.5$,
a cold probe with $\chi\le16$ and at most ten sweeps tests for an
energy improvement exceeding $10^{-6}\max(1,|E|)$. If successful,
the state is refined at the solver's full bond cap and accepted only
if it retains that improvement. Sites follow the input variable order.

A degenerate optimum may be represented by a superposition, so
single-site expectation values need not identify an optimal bitstring.
The readout therefore considers eight direct MPS samples, the sign
vector of $\langle Z_i\rangle$ and its global flip, and four random
resolutions of sites with near-zero expectation values. Each
candidate is refined by single-flip local search and the lowest
energy is returned.
The near-zero condition is $|m_i|<0.25(\max_j|m_j|+10^{-12})$,
where $m_i$ is the magnetisation returned by the MPS library.
Sampling, random resolutions, and polishing use the instance seed
in the primary single-anneal run.

\paragraph{LR-QAOA.}
LR-QAOA uses the fixed linear schedules
$\gamma_k=(k/p)\Delta$ and $\beta_k=(1-k/p)\Delta$,
$k=1,\ldots,p$, on the coefficient-normalised
Hamiltonian~\cite{MontanezBarrera2025}. Here $p$ is the layer count
and $\Delta$ is the schedule amplitude. The nonconstant Ising coefficients
are divided by $\max(\max_{i<j}|J_{ij}|,\max_i|h_i|,10^{-12})$.
Each layer applies the cost unitary followed by the mixer
$\exp(+i\beta_k\sum_iX_i)$, starting from $\lvert+\rangle^{\otimes n}$.
The Campaign A configurations and eligibility limits are given in
Table~\ref{tab:budgets}.
Integer division of the nominal 4000-shot budget gives 666 samples
per setting, or 3996 in total. The MPS runs use a truncation cutoff
of $10^{-10}$. The 3000-gate guard counts logical
two-qubit cost gates ($pm$ here), excluding single-qubit gates and
routing overhead. The observed data records 75 instances as SA fallbacks.
Their exclusion is described in Sec.~\ref{sec:solverdesc}.
Supplementary Campaign D protocols are specified in
App.~\ref{app:circuitsupport} and Sec.~\ref{sec:probscaling}.

The MPS implementation of LR-QAOA reproduces the statevector
implementation at $n=10$ with ample bond dimension, checking gate
conventions and the mixer sign. Larger-instance convergence is assessed
separately in App.~\ref{app:circuitsupport}.

\section{Supporting experiments and exploratory quantum diagnostics}
\label{app:supporting}

\subsection{Spectral protocols and capacity-failure breakdown}
\label{app:tnsupport}

The spectral routines use sparse Lanczos diagonalisation with at most
5000 iterations and no explicit override of the eigensolver tolerance.
The gap-versus-failure study uses 15 points from $s=0.3$ to $0.97$
on 3-regular instances at $n=14,18$, with seeds $0,1,2$, and six
anneals per instance. It tested whether a small minimum $\Delta_g$
predicts trapping. The pre-registered verdict was negative. With four
instances showing trapping and two without, the sample is below the
AUC reporting threshold of Sec.~\ref{sec:metrics}, so we give the
pattern instead. At an
essentially fixed minimum $\Delta_g\approx0.28$--$0.30$, the trap rate
rose from at most $1/3$ at $n=14$ to at least $5/6$ at $n=18$. Two of
the six instances have endpoint degeneracy $g=12$ and $18$; for them
$\Delta_g$ measures the distance to the 13th or 19th level, not an
excitation gap at intermediate $s$, so their values ($0.63$, $0.69$)
are not comparable with the $g=2$ cases. The negative verdict is
therefore weak evidence; the trapping account in
Sec.~\ref{sec:tnmitigation} rests on the trajectory comparison of
Fig.~\ref{fig:trapping}, not on gap magnitudes.
The trapping figure instead evaluates four low-lying levels on its
17-point solver grid, using exact diagonal energies at $s=1$.

Comparing $\chi=4$ and $16$ across all $90$ out-of-sample instances,
six failures at both caps are flagged as trapped, one more is cured at $\chi = 16$, and
one fails only at that larger cap. This non-nested pattern points to
residual trapping rather than a simple capacity limit. Accordingly,
$\chimax$ does not meet the pre-registered AUC target of $0.9$ as a
capacity-failure predictor (with capacity failure as the positive class
and larger $\chimax$ as the higher-risk score, AUC $0.88$ in sample and
$0.74$ out of
sample): capacity failures are too rare for reliable discrimination.

\subsection{Supplementary LR-QAOA coverage and convergence checks}
\label{app:circuitsupport}

The adaptive-depth experiment retains $\chi=32$, $\Delta=0.5$, and
500 samples, with $p=\min(64,\lfloor3000/m\rfloor)$. It covers all
27 main-benchmark instances with actual $n=32$. The corresponding grid
instances have $n=35$ and are excluded. These runs supplement, rather
than replace, the Campaign A fixed-depth comparison.

The convergence probes use six gauge-randomised Wishart instances at
$n=40,48$, with gauge seeds different from those of Campaign B.
Eighteen usable runs compare $\chi=8,16,32$ at $p=8$. One additional
usable run uses $p=40$, $\chi=32$ at $n=40$. Two additional attempts
terminate with singular-value-decomposition (SVD) errors and provide no convergence evidence.
The coverage and convergence outcomes are reported in
Sec.~\ref{sec:circuitchecks}.

\subsection{Exploratory candidate indicator and conditional resource projection}
\label{app:quantumexploration}

These analyses use main-benchmark diagnostics and a conditional
quantum-circuit resource calculation. They are not additional solver campaigns or
validated rules for predicting quantum advantage.

\begin{figure}[tbp]
  \includegraphics[width=\linewidth]{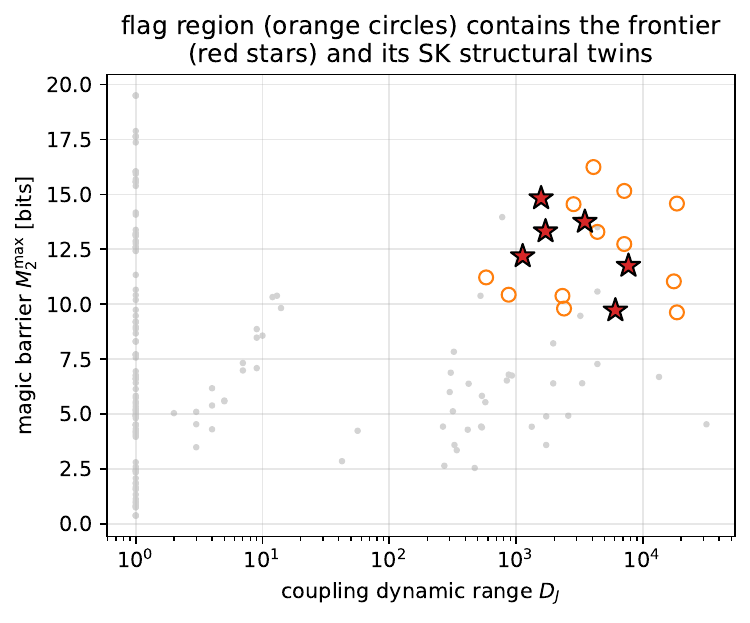}
  \caption{The quantum-candidate indicator in the $D_J$--$\Mmax$ plane. The
  selected region (orange circles) captures every Campaign A instance where
  SA fails against the reference optimum (red stars, all
  Wishart at $n \ge 32$) at in-sample recall $1.0$, but also twelve instances on
  which SA succeeds ($9$ Sherrington--Kirkpatrick
  structural twins and $3$ Wishart), indistinguishable by static features
  (precision $0.33$). Grey dots are the remaining instances of the
  Campaign A benchmark: not selected by the indicator, and solved by simulated
  annealing.}
  \label{fig:flagfrontier}
\end{figure}

We use the structural indicator
\begin{equation}
  \mathcal{Q}(Q) = \underbrace{[\rho_G \ge 0.9]}_{\text{TN priced out}}
  \wedge \underbrace{[D_J \ge 5] \wedge [h_{\mathrm{LM}} \le 0.3]}
        _{\text{wide-range, rugged}}
  \wedge \underbrace{[\Mmax \ge 8\ \text{bits}]}
        _{\text{no Clifford shortcut}},
  \label{eq:flag}
\end{equation}
to mark candidate instances where the measured classical routes are
strained. Its components combine dense-graph/treewidth structure, a
wide coupling range with a rugged local-minimum landscape (a static
profile shared by the Wishart, SK and random-frustrated ensembles, not a
measured predictor of SA failure), and a magic barrier that excludes
stabilizer-based simulation. All four thresholds were chosen on the
Campaign A data, against the six SA misses that the indicator is then
scored on. Its perfect recall is therefore an in-sample property, not a
validated detection rate. Precision is $0.33$ (Fig.~\ref{fig:flagfrontier}): nine false positives are SK instances and
three are Wishart instances on which SA succeeds. These structural twins
show why $\mathcal{Q}$ marks a candidate region rather than predicting
failure per instance. It is a separate diagnostic, not a runtime-based
solver recommendation.

The asymptotic argument is conditional. On all-to-all connectivity,
the $R_{ZZ}$ gates can be scheduled in
$\chi_{\mathrm{edge}}(G)$ rounds, the edge-chromatic number of $G$.
By Vizing's theorem~\cite{Vizing1964}, this is at most
$d_{\max}(G)+1$, where $d_{\max}(G)$ is the maximum vertex degree.
Including one mixer round per LR-QAOA layer, the gate-level circuit
depth is
\begin{equation}
  D = 1 + p\,[\chi_{\mathrm{edge}}(G)+1]
  \le 1 + p\,[d_{\max}(G)+2],
  \label{eq:depth}
\end{equation}
with $pm$ two-qubit gates. The literature reports near-constant success
probability at $p\sim n$ for several families~\cite{MontanezBarrera2025}.
If this carries over to Wishart, dense-instance depth grows as
$D\sim n^2$. The relevant classical comparison would then be exact
contraction, whose cost grows as $\exp[O(\mathrm{tw})]$ with
$\mathrm{tw}\sim n$ on dense graphs, not the bond-capped DMRG anneal of
this paper, whose cost is polynomial in $n$ at fixed $\chi$ and which
carries no optimality guarantee. That premise is not established here.

\section{Cost models and limits of heuristic selection}
\label{app:costmodels}

This appendix documents the selection procedure evaluated in
Sec.~\ref{sec:advisor}, as implemented in the accompanying code. It has
two stages. The first asks whether cheap Tier-0 features already
guarantee that an exact method is affordable. If so, there is nothing
to select: the instance is solved exactly, with a proof, and no
heuristic is considered. Only instances that this \emph{exactness check}
leaves unsettled reach the second stage, which ranks the heuristics by
predicted runtime. The distinction matters for reading the results:
the retrospective cost analysis below also covers instances that the
first stage would settle, so it evaluates more instances than the
procedure would actually rank.

The deployed exactness check applies four sufficient conditions in order
(Table~\ref{tab:stagea}). If one applies, it returns an exact method and
no heuristic cost is estimated. The cost fits below are a retrospective
analysis of observed timings on the calibration set, including instances
that the deployed $n \le 30$ rule would settle before cost ranking. In
deployment, these cost laws apply only after the exactness check leaves an
instance unsettled.

\begin{table}[tbp]
\caption{The exactness check: four sufficient conditions for an exact
classical method to be affordable, tested in this order on Tier-0 features
alone. The thresholds are the deployed defaults ($n \le 30$,
$\widehat{\mathrm{tw}} \le 25$). When one fires the instance is settled
exactly, the method is returned with the cited theorem as its explanation,
and no cost law is evaluated. R1's threshold is the pipeline's
default guard. Campaign A cross-checks with
enumeration only to $n \le 26$ (Sec.~\ref{sec:solverdesc}).}
\label{tab:stagea}
\footnotesize
\begin{tabularx}{\textwidth}{@{}lYYY@{}}
 & condition on $Q$ & why an exact method is then affordable & what is returned \\
\midrule
R1 & $n \le 30$ & all $2^n$ assignments can be enumerated in Gray-code order with $O(n)$ work per single-bit step, $O(n 2^n)$ in total; in our implementation ${\approx}\,25$\,s at $n = 24$, extrapolating to roughly half an hour at $n = 30$ & certified enumeration: the optimum \emph{with} a proof \\
R2 & planar graph, no fields ($h = 0$) & planar Max-Cut reduces to a matching problem and is solvable in polynomial time~\cite{Hadlock1975} & the polynomial exact algorithm \\
R3 & $\widehat{\mathrm{tw}} \le 25$ & exact contraction, equivalently dynamic programming over a tree decomposition, costs $\exp[O(\mathrm{tw})]\,\mathrm{poly}(n)$~\cite{MarkovShi2008} & exact contraction; no heuristic needed \\
R4 & balanced signed graph & no frustrated cycle, so the instance is switching-equivalent to a ferromagnet~\cite{Harary1953} & for $h = 0$, the ground state read off the switching 2-colouring in $O(m)$; single-flip descent alone is not exact, as it can stop at domain walls \\
\bottomrule
\end{tabularx}
\end{table}

For the retrospective cost analysis, we fit cost-to-target models for
simulated annealing (SA), brute-force enumeration (BF), and the
matrix-product-state tensor-network (TN) anneal. The predicted runtime is
\begin{align}
  \hat t_{\mathrm{SA}} &= \kappa_{\mathrm{SA}} R n_{\mathrm{sw}} n,
  \label{eq:tsa}\\
  \hat t_{\mathrm{BF}} &= \kappa_{\mathrm{BF}} n 2^n,
  \label{eq:tbf}\\
  \hat t_{\mathrm{TN}} &= \kappa_{\mathrm{TN}} n_s n \hat\chi^3,
  \qquad \log_2 \hat\chi = a + b\,\widehat{\mathrm{tw}},
  \label{eq:ttn}
\end{align}
where $R=4$ restarts, $n_{\mathrm{sw}}=2000$ sweeps, and the nominal TN
model uses $n_s=17$ grid points. This is a fitted runtime-model constant,
distinct from the solver's 11-point schedule; only the product
$\kappa_{\mathrm{TN}}n_s$ is identifiable. Two-site DMRG contributes the
$\chi^3$ scaling. Constants are fitted on the 90 calibration instances
using timings only. The treewidth regression predicts bond dimension from
Tier-0 structure; when Tier-1 data are available, measured $\chi_{99}$
replaces $\hat\chi$. SA, BF, and TN therefore have different degrees of
instance dependence. No cost law is fitted for PT, which is budgeted
directly in wall-clock time.

We evaluate predictions by the symmetric fold error
$f=\max(\hat t/t,t/\hat t)$. SA has median errors of 1.10 in sample and
1.22 out of sample, largely reflecting its fixed budget. BF has a median
error of 1.28 on the calibration set only. It is not run at $n\ge32$. For
TN, the out-of-sample median error is 7.59 using treewidth alone and 1.78
using measured $\chi_{99}$. The corresponding 90th percentiles are 189
and 36. Thus Tier-0 structure gives a coarse TN cost estimate, while the
smaller error requires a trajectory diagnostic.

\begin{figure}[tbp]
  \includegraphics[width=\linewidth]{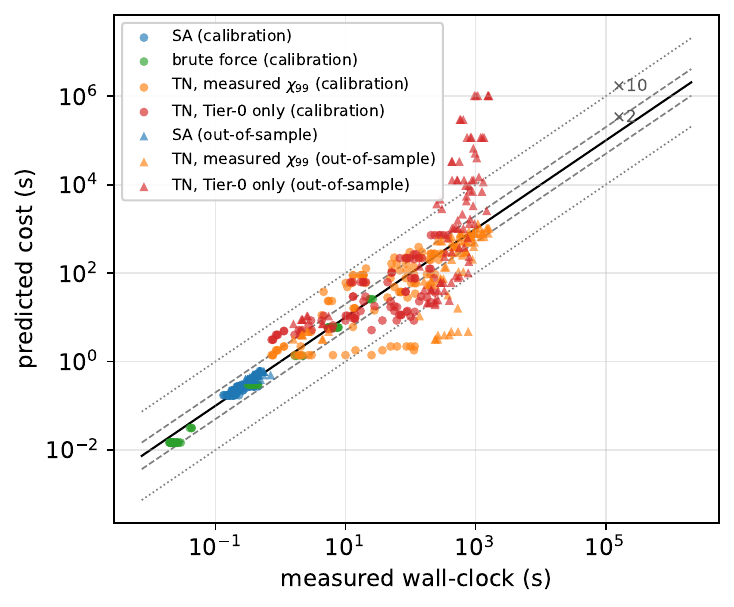}
  \caption{Predicted cost against measured wall-clock for the three cost
  laws, Eqs.~\eqref{eq:tsa}--\eqref{eq:ttn}, on both splits (circles:
  calibration, $n = 14$--$24$; triangles: out-of-sample, $n = 32$--$48$).
  The solid line is equality, the dashed and dotted lines factors of $2$
  and $10$. Simulated annealing and brute force track their own fixed
  budgets and sit on the diagonal. The tensor-network law scatters, and
  scatters much further when the bond dimension is inferred from the
  treewidth bound (Tier-0 only) instead of measured. Its largest
  out-of-sample predictions then exceed the measured times by three orders
  of magnitude. Brute force appears on the calibration split only: the
  enumerator is not run at $n \ge 32$.}
  \label{fig:costacc}
\end{figure}

Figure~\ref{fig:costacc} summarises the calibration and out-of-sample
errors for the three fitted cost laws.

The cost ranking $P^*(Q)=\arg\min_P\hat t_P(Q)$ is evaluated against
the ex-post heuristic labels in Sec.~\ref{sec:advisor}.

Under the fitted nominal budgets, the TN-to-SA predicted runtime ratio is
approximately $1.008\hat\chi^3$, so this benchmark's ranking favours SA
regardless of instance. This reflects the compared budgets and fitted
constants, not a general result about the paradigms. Weighting runtime by
a predicted hit probability also failed to improve the ranking. The
calibration set has no failures, so a model fitted there predicts unit hit
probability and leaves the ranking unchanged. An exploratory fit using all
instances likewise left the 85/90 score unchanged. A hard probability
cutoff reduced it. Together with the Wishart--SK feature twins, this
indicates that static features in this benchmark do not supply the quality
signal needed to repair the Campaign A ranking.
\end{appendices}

\bibliography{refs}

\end{document}